\documentclass[final,3p,times,authoryear]{elsarticle}
\usepackage{amssymb}
\usepackage{url}
\usepackage{amsthm}
\usepackage[utf8]{inputenc}
\usepackage{textcomp}
\usepackage{newunicodechar}
\newunicodechar{−}{\textminus}
\usepackage{amsmath}
\usepackage{booktabs}
\usepackage[hidelinks]{hyperref}

\newtheorem{resq}{Research Question}

\usepackage{caption}
\usepackage{soul}
\usepackage{xcolor}
\usepackage{overpic}

\newcommand{\msmc}[1]{\textcolor{blue}{msm says: \textit{#1}}}

\journal{---}

\usepackage{comment}

\usepackage{xcolor}
\definecolor{royalpurple}{rgb}{0.471, 0.318, 0.663}

\usepackage{soul}

\usepackage{ulem}
\begin{document}

\begin{frontmatter}



\title{From macro to micro: Economic complexity indicators for firm growth}

\author[inst1,inst2]{Maddalena Mula\corref{cor1}}
%\ead{maddalena.mula@cref.it}
\affiliation[inst1]{organization={Department of Economics and Finance},%Department and Organization
            addressline={Tor Vergata University of Rome}, 
            city={Rome},
            postcode={I-00133}, 
            country={Italy}}
\affiliation[inst2]{organization={Enrico Fermi Research Centre},
            addressline={Via Panisperna 89 A}, 
            city={Rome},
            postcode={I-00184}, 
            country={Italy}}
\cortext[cor1]{Corresponding author}

\author[inst3]{Valerio De Stefano}
%\ead{valedst97@gmail.com}
\affiliation[inst3]{organization={Department of Engineering and Science},%Department and Organization
            addressline={Universitas Mercatorum}, 
            city={Rome},
            postcode={I-00186}, 
            country={Italy}}

\author[inst4,inst5]{Manuel Sebastian Mariani}
\affiliation[inst4]{organization={Department of Business Administration},
            addressline={University of Zurich}, 
            city={Zurich},
            postcode={CH-8032},
            country={Switzerland}}
\affiliation[inst5]{organization={Department of Network Science, Institute of Data Analytics and Information Systems},
            addressline={University of Budapest}, 
            city={Budapest},
            postcode={HU-1093},
            country={Hungary}}

\author[inst6,inst2]{Andrea Zaccaria}
\affiliation[inst6]{organization={Istituto dei Sistemi Complessi (ISC-CNR)},
            addressline={UoS Sapienza},
            city={Rome},
            postcode={I-00185},
            country={Italy}}

% % use optional labels to link authors explicitly to addresses:

\begin{abstract}
A rich theoretical and empirical literature has investigated the link between export diversification and firm performance. 
Prior theoretical works hinted at the key role of capability accumulation in shaping production activities and performance, without however producing product-level indicators able to forecast corporate growth.
Building on economic complexity theory and the corporate growth literature, this paper examines which characteristics of a firm’s export basket anticipate future performance. We analyze a unique longitudinal dataset that covers export and financial data for 57,586 Italian firms. We find that firms exporting products typically exported by wealthier countries -- a proxy for products
characterized by a greater level of productivity and sophistication -- tend to experience higher growth and labour productivity. Crucially, where firms diversify matters: Across firms, only diversification outside of a firm’s core export area is positively associated with future growth, whereas diversification within the core is negatively associated. This is revealed by introducing novel measures of in-block and out-of-block export diversification, based on algorithmically-detected blocks. Our findings suggest that growth is associated not just with how many products a firm exports, but also with where these products lie within the production ecosystem, at both local and global scales. 
\end{abstract}


\end{frontmatter}

\section{Introduction}\label{sec:intro}

Indicators based on export baskets have been consistently linked with growth at the national level, yet their ability to predict firm-level growth remains contested.
At the macro level, country-level diversification-related indicators are a marker of a country's development stage~\citep{imbs2003stages} and they have been used to forecast future economic development~\citep{hausmann2007,tacchella2018dynamical,cadot2011export}. 
According to the Economic Complexity literature, this empirical evidence reflects a capability accumulation process where countries gradually acquire the know-how required to produce increasingly sophisticated goods~\citep{hausmann2011network}. As a result, the most complex products tend to be exported only by the most advanced and diversified economies -- a pattern reflected in the nested structure of global trade networks~\citep{tacchella2012new,mariani2019nestedness}.
However, translating this empirical evidence from the macro level (countries) to the micro level (firms) remains elusive, as the large body of literature on the diversification-growth link has produced mixed empirical evidence~\citep{dosi2022firm}. Hence, we address here a fundamental open question: Which characteristics of a firm's exported products anticipate firm growth?

Answering this question is not trivial. While classical perspectives emphasize the benefits of diversification and capability accumulation~\citep{penrose1959,teece_towards_1982}, empirical evidence on the effect of diversification for firm performance is inconclusive (see Section \ref{sec:lit}). 
That diversification alone cannot fully anticipate growth dynamics is also echoed by the observation that firms often possess a broader technological base than what is reflected in their actual output~\citep{PATEL1997,bottazzisecchi2006,dosi2017}, indicating a high degree of selectivity in product development. Not all firms aim to diversify as widely as possible; instead, many strategically concentrate on coherent or closely related activities aligned with their existing capabilities~\citep{teece1994understanding,dosi2022firm}, which results in a highly-modular structure of firms' exporting patterns~\citep{laudati2023different}. 
While coherent diversification -- diversification into areas related to existing capabilities -- has been linked to higher labor productivity ~\citep{pugliese2019coherent}, recent evidence found no association between coherence and future firm growth~\citep{dosi2022firm}.
Overall, identifying firm-level indicators of diversification and export basket characteristics that anticipate future growth remains an open and unresolved problem.

To address this open challenge, we leverage here a unique dataset made available through a collaboration with the Italian National Institute of Statistics (ISTAT). It includes detailed firm-level export records, with products classified using the same scheme (Harmonized System) employed in country-level trade data. We integrate this with firm-level financial information, covering over $50,000$ Italian firms across multiple years. This rare combination enables testing economic complexity theories and metrics previously developed through countries' export data. 
Crucially, the combination of high-resolution export data and financial data allows us to apply network analysis techniques to identify each firm’s core export block -- a cluster in the firm-product bipartite network identified via modularity optimization~\citep{barber2007}. 
This enables us to distinguish between diversification within and outside the firm's core block, and to correlate the resulting indicators with both firm growth and labour productivity.

Building on the existing literature and this unique dataset, we explore how the characteristics and structure of a firm’s export basket relate to future firm performance. Our investigation centers on two main dimensions. First, we ask whether for a firm, exporting products that are exported by wealthier countries is positively associated with future growth.
This builds on the idea that the sophistication of a firm's product basket reflects the strength of its underlying capability base~\citep{hausmann2011network}. It mirrors findings at the country level, where exporting more complex products signals capability accumulation and predicts economic development~\citep{hidalgo2009}.
To answer this question, we adapt the country-level indicator EXPY~\citep{hausmann2007} for firms, and demonstrate that it exhibits a positive association with both firms’ future growth and labour productivity.

\begin{figure}[t]
    \centering
    \includegraphics[width=.8\linewidth]{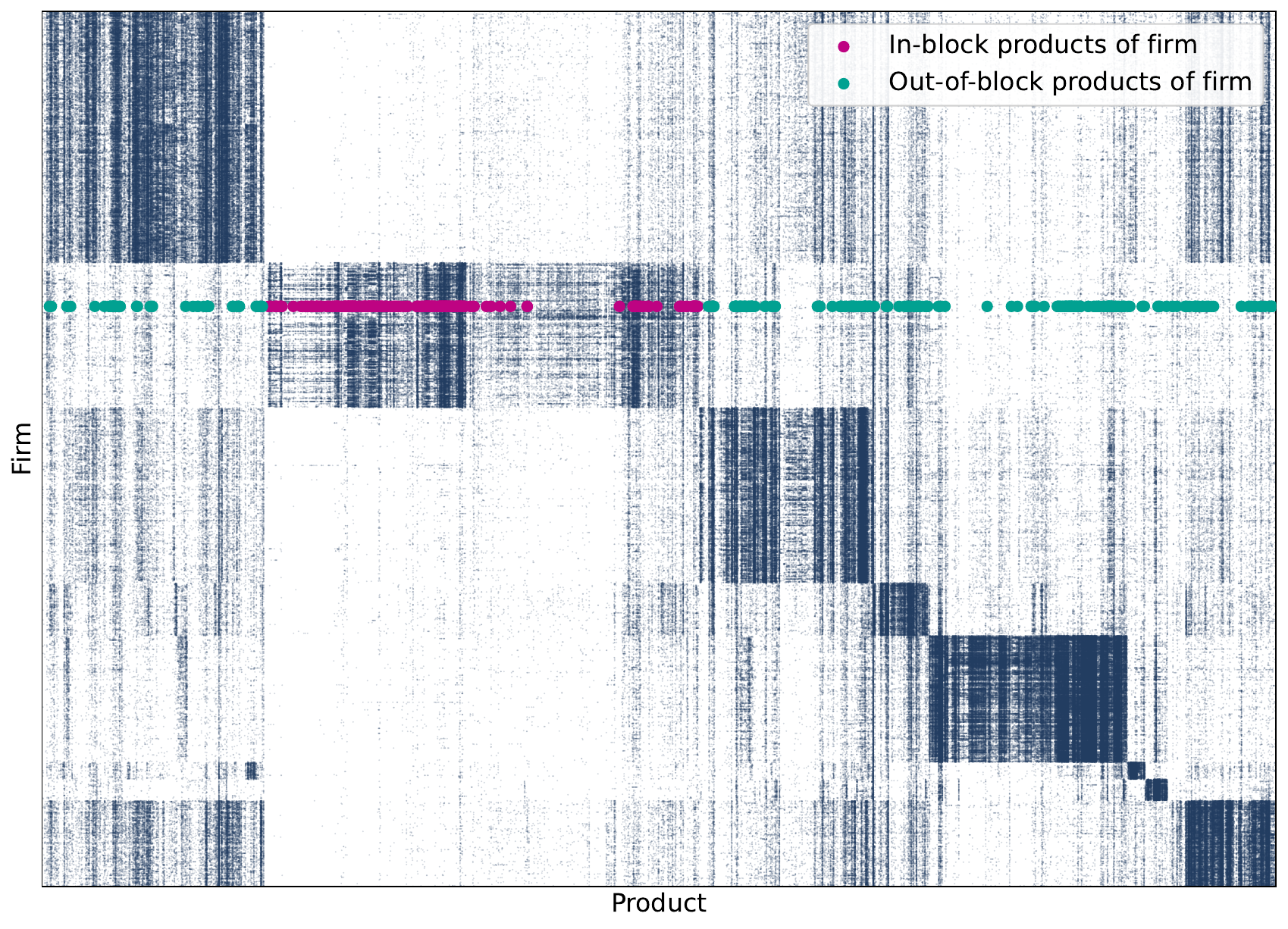}
    \caption{\textbf{An illustration of the modular structure of the firm-product export network.} The binary matrix's elements represent whether a given firm (row) exports a given product (column) above a certain threshold. We used a bipartite modularity optimization algorithm to partition the matrix into blocks (for the data definition, see Section~\ref{sec:data}; for a detailed description of the algorithm, see \ref{brim_section}). For a given firm, we can distinguish among exported products within the firm's block (purple dots) and outside the block (green dots). This algorithmic identification of firms' core production areas allows for distinguishing between in-block and out-of-block diversification.} 
    \label{fig_matrix} 
\end{figure}

Furthermore, inspired by macroeconomic findings that emphasize the benefits of broad diversification~\citep{hausmann2011network,metcalfe2002evolutionary,saviotti2008trade}, we ask whether and how diversification is beneficial at the firm level. This question has a long tradition in the strategic management literature, but has seen conflicting empirical answers (see Section~\ref{sec:rq}). Here, we investigate whether the structure of diversification -- namely, whether it occurs within or outside a firm’s core export area -- matters for growth.
To explore this distinction, we apply a community detection algorithm to the firm–product export network to identify each firm's core production area, which we refer to as a ``block". We then introduce two novel diversification indicators (the in-block and the out-of-block diversification) that separately capture expansion within and beyond these core product clusters (see Fig.~\ref{fig_matrix} for an illustration).
Surprisingly, our results reveal a stark contrast: across firms, diversification within the core (in-block) is negatively associated with firm growth, while diversification beyond the core (out-of-block) is positively associated.
Turning from differences across firms to changes within firms, our fixed-effects analysis reveals a significant negative association for in-block diversification across all time horizons, whereas the association for out-of-block diversification is sensitive to the treatment of outliers (see Section \ref{panel_regression}).
Together, these findings suggest that growth is associated not only with how much a firm diversifies, but also with its specific diversification strategies with respect to its core export area.

Our work offers two major contributions to the two research lines on economic complexity and firm growth.
\textit{First}, it develops methods of economic complexity theory, traditionally applied at the country level, for the firm level, thereby bridging macroeconomic capability-based approaches with micro-level analyses of corporate performance. In doing so, it clarifies which economic complexity indicators are associated with firm-level outcomes such as growth and labour productivity. This opens up new possibilities for future research on firm dynamics using tools from the economic complexity framework.
\textit{Second}, the study contributes to long-standing debates on the relationship between diversification and firm growth. While diversification does matter, its effect is subtle. Across firms, diversification outside the core is positively associated with subsequent growth, whereas diversification within the core exhibits a negative association. Departing from the established literature on corporate coherence, this result becomes visible only through network-based analyses grounded in economic complexity. Finally, this finding highlights the value of looking beyond aggregate diversification metrics and toward the structural position of new activities within a firm’s broader capability space.

\section{Background}\label{sec:lit}

The ability of a firm to diversify its production and the markets in which it operates is widely recognized as a key driver of long-term growth and performance. Diversifying production involves expanding the range of products a company develops to reach new customer segments within markets other than those it already operates in \citep{metcalfe2002evolutionary}. In her seminal work "The Theory of the Growth of the Firm", \cite{penrose1959} argued that firms grow by expanding their range of production, a process inherently linked to diversification. This diversification enables firms to take advantage of new market opportunities, mitigate risks associated with individual sectors or products, and improve their overall resilience~\citep{teece_towards_1982}. 

This approach follows from a research strand, called \textit{resource-advantage theory}, which views firms as a collection of resources and \textit{capabilities} and explores the importance of coherence within such a bundle~\citep{winternelson1982,dosi1988sources}. According to \cite{sutton2001}, the term \textit{capability} relates to the set of ‘shared know-how’ embedded within a firm's workforce, which underpins its levels of productivity and quality. In a dynamic environment, where demand and supply conditions fluctuate across multiple markets, a firm's growth depends not only on its current productivity and quality within its product bundle but also on the underlying know-how that will determine its potential performance in new product groups it may enter.
\cite{BellPavitt1993} introduced the concept of dynamic capabilities of the firm, which was further developed by \cite{teece1994}, defining dynamic capability as the ability of a firm to integrate, build, and reconfigure internal and external capabilities to address rapidly changing markets. \cite{teece1994} elaborate on how firms can take advantage of their set of dynamic capabilities to enter new markets and produce new products, ensuring sustained growth. Subsequent work has reinforced this perspective, demonstrating that firms with strong dynamic capabilities exhibit greater adaptability and resilience in volatile environments ~\citep{eisenhardt2000}.
Note that, in principle, the organizational/dynamical capability theory supports both the tendency of firms to diversify in related activities, as in \cite{pugliese2019coherent,dosi2022firm,bontadini2023being}, and their ability to explore new opportunities beyond the production core~\citep{matsusaka2001corporate,bernardo2002resources}.

Technological homogeneity in firms' capabilities can coexist with product variety. As argued by \cite{PATEL1997}, the technological capabilities of large firms are typically multi-field and more diversified, often extending beyond their actual product range. Firms can maintain technological capabilities in multiple domains with a knowledge base that is larger than what their output would suggest, which has also been observed for large patentees among Italian firms~\citep{dosi2017}. Specifically, \cite{bottazzisecchi2006} proved that there exists an exponential relationship between the size of a company and the number of products exported. In general, larger firms benefit from broader capabilities, which provide a wider base for potential product diversification. In contrast, smaller firms may struggle to allocate resources for diversification, making them more vulnerable to market shocks. 

To better understand the internal structure of firms’ production and technological activities, recent research has formalized the concept of \textit{coherence}, which captures the degree of relatedness among the elements in a firm’s product or technological portfolio. Coherence refers to the degree of relatedness between products in a firm's portfolio: it measures firm-level consistency in the capabilities required to produce different products ~\citep{teece1994}. Hence, coherent or related diversification occurs when a firm can diversify into areas that are related or connected to its core business. 
\cite{Piscitello2000} found that while the relatedness between firms’ products decreases over time, the relatedness between technologies increases, which indicates that firms maintain coherence in their technological capabilities even as they diversify into less related products. However, \cite{Piscitello2000} adopts a coarse-grained classification which considers about 50 industries (while in the present paper we consider more than 5000 different products) and does not establish a direct link between relatedness and firm performance or growth. \cite{barbieri2024export} investigate the relationship between geographical and sectoral diversification of exports and the response to global economic shocks.
More recently, \cite{pugliese2019coherent} measured technological relatedness within firms as co-occurrences of technology classes within patents and examined the relevance of coherent technological diversification in explaining firm performance. Their metric explains labor productivity above and beyond simple technological diversification.
Their findings are echoed by \cite{dosi2022firm}, who analyzed the relationship between diversification, coherence of the product basket, and firm performance, both in terms of growth and profitability. They found that firms that pursue coherent product diversification are more likely to experience improvements in labor productivity, but not in firm growth~\citep{dosi2022firm}.
Whether indicators based on the structure of a firm’s production portfolio can anticipate growth remains an open question, which we aim to address in this article. Moreover, previous literature \citep{march1991exploration,o2013organizational} distinguishes between the exploitation of existing capabilities and thus the building of a coherent basket of products, and the capability of exploring production areas which are, in principle, out of the main industrial strategy of the company and its competitors. In this paper, we will quantify these concepts by introducing in-block and out-of-block diversification strategies. 

Although firm-level studies highlight the role of coherence in supporting productivity, the macroeconomic literature on countries’ growth presents a contrasting logic: countries are expected to grow by acquiring new capabilities \citep{fagerberg2008national,fagerberg2010innovation} and diversifying into increasingly complex and unrelated products.
Countries grow by acquiring and combining capabilities to diversify their export~\citep{hidalgo2007, caldarelli2012, zaccaria2014}. This is reflected in the nested structure of the network that links countries to the products they export. At the country level, specialization in \textit{complex} exports is associated to growth: more complex products are linked to countries with higher growth rates and economic development~\citep{hidalgo2009, tacchella2018dynamical}. A product is complex if only a few highly diversified countries export it, meaning that its production requires rare capabilities that are only acquired by the most developed countries.
However, at the firm level, significant network differences emerge. As discussed above, firms tend to focus their production on coherent sets of products. This is further reflected in the structure of firm-product export networks, which link firms to exported products. \cite{laudati2023different} and \cite{bruno2018} show that, unlike country-product networks, firm-product networks exhibit a modular structure (see Figure \ref{fig_matrix}) 
where firms tend to export products in well-defined clusters of related products (i.e., production sectors). This structural coherence in diversification is based on the ability of firms to manage and combine different capabilities that are necessary to produce a huge variety of products, which agrees with a capability-based view of diversification.
These structural differences raise an important question: can the economic complexity methods developed for countries be meaningfully adapted to investigate firms' growth? In this paper, we explore this possibility.

A strand of research in economic geography has examined the relationship between related variety, unrelated variety, and economic growth at the regional level (\cite{frenken2007related,neffke2011regions,boschma2009related}). These studies compute measures of affinity, or relatedness, among the different sectors represented in a region’s industrial production structure (\cite{balland2017geography}). Such measures are then aggregated at the regional level in a way analogous to the concept of coherence discussed above for firms, and related to various indicators of growth. The main finding is that a coherent technological and productive portfolio, or one characterized by related variety, is more frequently associated with employment growth at the regional level (\cite{frenken2007related}). Together with measures of economic complexity, these approaches provide a framework for so-called smart specialization strategies (\cite{balland2019smart}) and for the analysis of regional economic development (\cite{content2016related}).

Our theoretical framework is grounded on the capability theory of firms; however, various scholars have investigated the relationship between export diversification strategies and firm performance using different perspectives.
\cite{mayer2014market} discuss the relationship between competition and the choice of which products to export. \cite{eaton2011anatomy} study the sales distributions of French manufacturing firms towards various destinations. \cite{melitz2003impact} theoretically shows the association between entering the export market and the productivity of firms. \cite{melitz2012gains} review different sources of gains from trade, including innovation and love-of-variety. According to \cite{eckel2015multi}, theoretical and empirical evidence shows that core products have higher quality, while firms' export strategies can be based on quality or cost. \cite{cirera2015explaining} investigates the firm-level determinants of export diversification, showing that the structural characteristics of the firms cannot account for all the observed heterogeneity.
None of these contributions correlates the characteristics of exported products with growth, which is the key feature of the economic complexity contribution to macroeconomics.

\section{Research questions}
\label{sec:rq}

To examine the relationship between export structure and firm growth, we draw inspiration from network science and the country-level literature on economic complexity, adapting its indicators to the firm scale.
This body of work has shown that measures of product sophistication and diversification can effectively predict national economic growth~\citep{tacchella2018dynamical,hidalgo2021economic,koch2021economic,sbardella2018role,angelini2024forecasting}.
However, the modular structure of the firm-product networks \citep{bruno2018, laudati2023different} suggests that the firms exhibit product diversification patterns that do not necessarily align with the country-level complexity metric. Measures of product complexity have been employed to investigate firm performance in terms of volatility \citep{maggioni2016} and total factor productivity \citep{guan2020}. \cite{becnkovskis2024aim} show that exporting complex products improves the survival of trade relationships while, interestingly, exporting products that are more complex or distant from the firm’s previous export basket reduces the chances of survival. Note that these contributions do not investigate complexity measures as direct predictors of firm growth. One possible limitation of complexity-based metrics is that they primarily capture how \textit{difficult} a product is to produce, rather than how \textit{economically valuable} it is \citep{hidalgo2009, tacchella2012new}. 

Consequently, we follow here an alternative approach to predict growth, based on the productivity level of the products a firm exports. We adopt a measure of the productivity level of firms' product basket, the PRODY index, which measures the level of income or productivity associated with a given product~\citep{hausmann2007}. Specifically, \cite{hausmann2007} introduced the level of productivity associated with the export basket of the country $c$, the index $\mathrm{EXPY_c}$, and showed that countries exporting goods associated with higher levels of EXPY tend to experience subsequent economic growth. As \cite{hausmann2007} argues: "Everything else being the same, an economy is better off producing goods that richer countries export". The underlying logic is that products exported by high-income countries \textit{(i)} tend to require more advanced capabilities -- whether technological, organizational, or human capital related -- and thus serve as indirect indicators of productive know-how~\citep{hausmann2011network,cristelli2013measuring}, and \textit{(ii)} are worthwhile and advantageous from a market perspective. 
Note that \textit{(ii)} implies that products need not be “complex” to score high on EXPY: a low-complexity product that is strongly demanded in international trade can still be exported heavily by high-income countries. Accordingly, EXPY is interpreted as a productivity measure~\citep{hausmann2007}, which is conceptually distinct from ``complexity" indices (e.g., fitness or ECI) that proxy the amount of capabilities required to produce and export a product~\citep{hidalgo2009,tacchella2012new}.
Adapting this reasoning to the firm level, firms exporting products associated with higher-income countries are likely to possess more complex and valuable capabilities and/or make strategic diversification choices, which may signal a stronger foundation for future growth and profit gains.

Building on this framework, we adapt the EXPY index to the firm level to assess whether firms exporting higher-productivity goods exhibit long-term growth. Consequently, we propose the following: 
\begin{resq}[RQ1]
Do firms that export products associated with higher levels of productivity tend to exhibit higher future growth and labour productivity?
\end{resq}
To answer this question, we test whether the indicator $\mathrm{EXPY_f}$, defined as the weighted average of $\mathrm{logPRODY_p}$ of all products exported by a company, is significantly and positively associated with Growth and labour productivity at the firm level.
In doing so, our objective is to identify a firm-level indicator inspired by the economic complexity framework that is significantly related to future firm growth, which is lacking in previous work.

In addition, our goal is to contribute to the ongoing debate on the relationship between diversification and firm growth by examining how growth outcomes depend not just on the extent of diversification but on its structure, specifically, how it relates to the block structure of the firm-product network.
As demonstrated by \cite{bruno2018,laudati2023different} and further discussed in this paper, firms tend to diversify their export within blocks of related products, which can be interpreted as the firm's core area of export and can be algorithmically identified. However, the diversification of firms is not limited to these clusters, as a significant share of their production is also linked to a broader structure that extends beyond related products, as found by \cite{dosi2017}. This suggests, as noted by \cite{PATEL1997} and \cite{Piscitello2000}, that firms operate within a dual framework. The first is an internally coherent one in which firms exploit shared capabilities within a cluster; we quantify this with a variable we call \textit{in-block diversification}. The second strategy is, instead, an externally expansive one (quantified by an \textit{out-of-block diversification}), and draws on exporting products that are not directly related to the primary block of production. 

The literature has highlighted that high levels of diversification outside a firm's core area may signal that the firm is underperforming in its current activities~\citep{lang1994tobin,gomes2004optimal} and undertaking high opportunity costs~\citep{maksimovic2002conglomerate}.
On the other hand, theoretical and empirical work has shown that diversification across sectors could reflect that firms possess capabilities that are transferable between products and industries~\citep{matsusaka2001corporate,bernardo2002resources}. Empirical evidence on the relation between diversification and performance has been mixed, with some studies pointing to a significant positive relationship~\citep{rumelt1982diversification,delios2008within,dosi2022firm}, but others pointing to a negative one~\citep{lang1994tobin,markides1995diversification} or even an insignificant one~\citep{christensen1981corporate}. On the other hand, \cite{bontadini2023being} showed that the performance of companies is related to the coreness, or centrality, of their products with respect to the respective export baskets.
This conflicting evidence leads us to use algorithmically-identified blocks to define different diversification strategies and formulate the following:

\begin{resq}[RQ2]
What is the relation between diversification \textit{outside} a firm's core export area and the firm's economic performance in terms of firm growth and labour productivity?
\end{resq}

Compared to diversification outside a firm’s core production area, product diversification \textit{within} a firm’s core area -- here referred to as in-block diversification -- remains less explored. While prior research has emphasized the role of coherence and internal relatedness in supporting profitability and resource efficiency~\citep{teece1994understanding,dosi2022firm,bontadini2023being}, it is less clear whether expanding further within an already coherent production space leads to better growth outcomes.
The macroeconomic literature on economic complexity suggests that even within a related product space, higher levels of diversification may signal richer capability endowments and be associated with stronger growth trajectories~\citep{hidalgo2007,hausmann2011network,boschma2015institutions}. Following this logic, in-block diversification could indicate a broad and coherent set of a firm’s core capabilities, which might signal future growth. However, organizational and learning theories raise a competing view: excessive reliance on related activities may reflect path dependence, bounded search, or lack of strategic renewal~\citep{march1991exploration}. In this view, in-block diversification might deliver limited marginal returns and be symptomatic of firms avoiding riskier (but potentially more productive) exploration outside their core.
To empirically assess whether in-block diversification supports or constrains firm performance, we formulate the following research question:

\begin{resq}[RQ3]
What is the relation between diversification \textit{inside} a firm's core export area and the firm's economic performance in terms of firm growth and labour productivity?
\end{resq}

To answer RQ2 and RQ3, we detect firms' ``block", \cite{laudati2023different}, through a community detection algorithm well established in network science (see \ref{brim_section} for details), and we measure the number of products that each firm exports within and outside its block.
Together, RQ2 and RQ3 aim to uncover whether the performance effects of diversification depend on where diversification occurs within a firm’s product space. If diversification outside the core supports growth, while in-block diversification does not, this would challenge conventional views on coherence by suggesting that strategic exploration beyond familiar domains may be more beneficial for long-term performance, according to the organizational learning literature~\citep{march1991exploration}. Answering these questions can provide firms with insights into the effects of strengthening their capability base within their core area as opposed to exploring unrelated capabilities.

\section{Data}\label{sec:data}

\subsection{Export data}

Our study is based on two primary datasets. The first dataset, CoeWeb (https://esploradati.istat.it/coeweb/databrowser), is provided by the Italian National Institute of Statistics (ISTAT) and contains export volume information for products exported by virtually all Italian firms. 
As we aim to establish the link between firms' export basket characteristics and growth, we restrict our analysis to firms that have consistently engaged in exporting activities over time. Specifically, we only retain firms that exported at least one product per year throughout the entire period covered by our dataset, from 2015 to 2021. This allows us to focus our study only on those firms that show a solid and established export strategy. Following this filtering process, $57,588$ 
%\textcolor{red}{queste sono le aziende dopo il filtro sugli export tutti gli anni, prima di controllare di quali abbiamo anche info finanziarie. aggiornare numero} 
firms remained. Export data specifies which company exported which product, in what year, and in what volume. Products are classified using Harmonized System codes with a six-digit resolution level, for a total of $5,389$ different products in our dataset.
This high level of product resolution enables a detailed mapping of the export portfolios of firms and is crucial to constructing our network-based indicators, particularly the in-block and out-of-block diversification measures. Moreover, Italy provides an ideal empirical context: as a country with high economic complexity and broad export diversification across sectors~\citep{tacchella2012new,sbardella2021behind}, it allows us to observe firm-level growth patterns across a wide range of industries. 
At the country level, export data have been previously shown to capture domestic production accurately, except for services~\citep{saltarelli2020export}. Future works could incorporate additional dimensions of firms' productive activity, such as domestic sales and technology portfolios, into the metrics' calculations, in order to provide a richer picture of firms' production activities.

\subsection{ASIA financial data}

%\textcolor{red}{rivedere i numeri delle aziende}
The second dataset was provided by ISTAT as well, and contains financial data for 
%a subset of the 18,597 firms in the CoeWeb dataset. Among the pre-filtered 18,597 
$331,510$ firms. The ASIA dataset includes financial information from 2015 onward for the $57,586$ firms. 
The available financial variables include the yearly number of employees, value added, labour productivity, age of the firm, geolocalization, economic sector. We refer to \url{www.istat.it/scheda-qualita/registro-statistico-delle-imprese-attive-asia-imprese-2/} for the detailed definition of the database and the included information. 
We aim to use the proposed economic complexity indicators to explain the trends of two types of performance of firms: labour productivity and growth. 
We computed the labour productivity as the ratio between the gross value added over the number of employees. We calculated the growth variable as the log difference between the four-year-ahead and current three-year moving averages of value added (see Section~\ref{sec:res}). 
%\textcolor{red}{maddalena controlla se le nuove definizioni sono giuste}

\section{Economic complexity indicators}

Motivated by the economic complexity literature, we introduce here two classes of economic complexity indicators. First, we adapt to firms the EXPY indicator originally introduced for nations~\citep{hausmann2007} (see Section~\ref{sec:expy}). Subsequently, building on a well-established modularity maximization algorithm~\citep{barber2007}, we introduce the in-block and out-of-block diversification indicators (see Section~\ref{sec:inoutblock}).
Finally, we define a coherence indicator closely related to that by \cite{dosi2022firm} (see Section~\ref{sec:coherence}).

\subsection{EXPY definition}
\label{sec:expy}

The goal of our EXPY metric is to measure the average productivity level of the products exported by a firm, a quantity which we aim to correlate with firms' economic performance.
To measure the productivity level of products, we adopt a slightly modified version of the PRODY index proposed by \cite{hausmann2007}. The PRODY index is calculated from the bipartite country-product network, represented by an export matrix whose element $E_{c,p}$ indicates the export volume of the product $p$ by country $c$. To factor out size effects, one does not look directly at the export volume, but at the Revealed Comparative Advantage \citep{balassa1965}, which is size-agnostic and defined as: 
\begin{equation}
    \text{RCA}_{cp} = \frac{E_{cp} / \sum_{p} E_{cp'}}{\sum_{c'}E_{c'p}/\sum_{c',p'}E_{c'p'}}
    \label{eq:rca}
\end{equation}
where the numerator captures the extent to which product $p$ is exported by country $c$, and the denominator captures the extent to which product $p$ is exported in the whole world. Therefore, RCA is a metric that represents how much a country exports a product both relatively to the rest of the world and relatively to its other exports of products.
Although the original definition of PRODY employs a weighted average of GDP among exporting countries, we use the logarithm of GDP instead. This choice reflects the highly skewed nature of the GDP distribution, which spans about four orders of magnitude \citep{Angelini2017}, and ensures a more balanced contribution of countries in the calculation of productivity at the product level.
The logPRODY of a product $p$ is defined as the weighted average of the logarithms of the GDP (per capita) of countries, where the weights are the corresponding RCA values \citep{Angelini2017}:
\begin{equation}
\label{prody_def}
    \text{logPRODY}_p = \frac{\sum_c \text{RCA}_{cp}\ln \text{GDP}_c}{\sum_c \text{RCA}_{cp}}.
\end{equation}
Equation \ref{prody_def} implies that products exported predominantly by high GDP (per capita) countries will exhibit a higher value of the logPRODY index, while products exported by low GDP (per capita) countries will exhibit a low score. 
The intuition behind this metric is that high-GDP countries are, on average, more productive than low-GDP ones, and one of the ingredients of their high productivity is given by the products they focus on. 

Even if the logPRODY metric is not explicitly connected with the concept of capabilities, we can, however, expect that it to contain some information about them, given its correlation with economic complexity measures of products' sophistication \citep{Angelini2017,laudati2023different}.
Indeed, the logPRODY index bears some resemblance to indicators of product complexity such as the Product Complexity Index (PCI)~\citep{hidalgo2009} and the Complexity derived through the fitness-complexity algorithm~\citep{tacchella2012new}.
The key difference between logPRODY and these product complexity metrics lies in the respective goals: the PCI and Complexity aim to capture the underlying capabilities required to export of a product, whereas the logPRODY captures the income levels of the countries that export the product. Although the two concepts are correlated\footnote{We find a statistically significant Spearman's correlation coefficient of $r=0.252$ ($P<0.001$) between PRODY and the product complexity indicator defined by \cite{tacchella2012new} and a statistically significant Spearman's correlation coefficient of $r=0.728$ ($P<0.001$) between PRODY and the product complexity indicator defined by \cite{hidalgo2009}.}, the product complexity indicators do not take into account that some products, while easy to produce, could still be associated with high productivity or be well received on the market. 
Various results in development economics indicate that, at the country level, exporting only low‐complexity products signals a lack of productive capabilities \citep{bebbington1999capitals,fagerberg2010innovation,pugliese2017complex,freire2019economic}. However, the same line of reasoning does not apply directly to firms' capabilities: for instance, firms exporting high-quality agrifood products may experience high growth even if the respective sector has low complexity. 

To improve the interpretability of the logPRODY index, we normalize it by subtracting its mean and dividing by its standard deviation. To simplify terminology, hereafter we will refer to the $z$ score of the original logPRODY index as logPRODY.
Based on the logPRODY scores of the products, we calculate the EXPY values for each firm as follows: 
\begin{equation}
\label{expy_def}
    \text{EXPY}_f = \frac{\sum_p E_{fp} \text{logPRODY}_p}{\sum_p E_{fp}},
\end{equation}
where $E_{fp}$ is the export volume of product $p$ associated with the firm $f$. Note that by using the export volumes as weights in Eq.~\eqref{expy_def}, our definition of the EXPY differs from that adopted by \cite{hausmann2007}, who used the RCAs as weights instead. The main reason for preferring to use the weights in the EXPY calculation is that in the RCA definition, the term $\sum_{c'}E_{c'p}$ in Eq.~\eqref{eq:rca} should conceptually capture the total global export volume of product $p$, which cannot be captured by only using data from Italian firms. However, results are robust with respect to both types of weights: export volume and RCA (calculated in the firm-product network).

\subsection{In-block and out-of-block diversification}
\label{sec:inoutblock}

To capture the extent to which a firm diversifies its exports across multiple production areas, here we introduce in-block and out-of-block diversification. 
The starting point is to partition the bipartite graph that connects firms and products into distinct blocks\footnote{Note that the term ``blocks" is meant to be more general than alternative terms widespread in the network science literature, such as ``modules" and ``communities"~\citep{fortunato2016community}.}, such that most of the products exported by firms within a given block also belong to the same block.
A standard method to achieve this objective is the maximization of bipartite modularity, for which several optimization algorithms have been introduced in network science~\citep{fortunato2016community}.
Here we apply the BRIM algorithm (Bipartite Recursively Induced Modules) because of its wide use, efficiency, and known reliability in bipartite graphs~\citep{barber2007}. We refer to \ref{brim_section} for definitions and implementation details.

By applying the BRIM algorithm to the export dataset, we find eight distinct blocks. Each block can be economically interpreted by examining the relative representation of the different industrial sectors (see Figure \ref{FigBlocksEco}). The largest block mainly includes machinery and metal products and represents about $30\%$ of the firms. Among the remaining seven blocks, five of them exhibit a clear majority of products in a specific sector: one block mostly features products in electronics (e.g., televisions, hard disks, etc.), one in paper and packaging, one in automotive and transport, one in textile products, and one in jewellery. The remaining blocks are less interpretable, as they include substantial proportions of products from the furniture and building materials industries.

Based on the detected blocks, we define a firm $f$'s in-block diversification as the number of products it exports that belong to the same block as the firm. Conversely, out-of-block diversification refers to the number of products exported by $f$ that belong to a different block from the one to which firm $f$ belongs (see Figure \ref{fig_matrix} for reference).
Formally, we can introduce the indicator variable $\delta_{fp}$ which is equal to one if firm $f$ and product $p$ belong to the same block, zero otherwise. 
We then define the binary matrix $\boldsymbol{M}$ as the binary matrix whose element $M_{fp}$ indicates whether firm $f$ exports a significant volume of product $p$. We use the Revealed Comparative Advantage\footnote{We used the Revealed Comparative Advantage because it allows for a size agnostic threshold. } (defined as in Equation \ref{eq:rca}) to determine the threshold: $M_{fp} = 1$ if $RCA_{fp}>1$, while $M_{fp} = 0$ if $RCA_{fp}\leq1$.  
Equipped with these variables, the in-block diversification and out-of-block diversification are simply defined, respectively, as
\begin{equation}
\begin{split}
    d^{in}_f &= \sum_p M_{fp}\delta_{fp},\\
    d^{out}_f &= \sum_p M_{fp}(1-\delta_{fp}).
    \end{split}
\end{equation}
The sum of these two variables is simply equal to the number of products that the firm is significantly exporting, which is the overall level of diversification of the firm: $d_f=d^{in}_f+d^{out}_f$.

\subsection{Coherence}
\label{sec:coherence}

Since the literature has found positive relationships between industrial and technological coherence and various forms of profit (see Section~\ref{sec:lit}), we include it as a control variable. 
Coherence metrics assess how closely related the products exported by a given firm are to each other. The concept of relatedness is computed using the similarity matrix $B$, whose element $B_{pp'}$ quantifies the similarity between the products $p$ and $p'$. While \cite{dosi2022firm} used product co-occurrences as a measure of similarity, here we adopt the Sapling Similarity, which has been shown to outperform alternative similarity measures in forecasting export activities~\citep{albora2023sapling} -- see \ref{sec:sapling} for details of the definition. 
We calculated firm-level coherence by weighting the similarity of the products based on their export volumes as follows:
\begin{equation}
    C_f = \frac{\sum_{p,p'}E_{fp}E_{fp'}B_{pp'}}{\sum_{p,p'} E_{fp}E_{fp'}},
\end{equation}
where $E_{fp}$ denotes the export volume of product $p$ by firm $f$. Therefore, a firm that exports only products with a high vicinity (that is, $B_{pp'}$ close to $1$) will have a coherence value close to $1$, while a firm that exports unrelated products (that is, $B_{pp'}$ close to zero) will have a Coherence value close to zero. In principle, since we are using the Sapling Similarity, negative values of Coherence are possible if enough couples with negative values of $B_{pp'}$ are present in a firm. However, there are no cases of negative Coherence in the data.

We point out that the Coherence measure discussed by \cite{dosi2022firm} differs in two aspects from the one introduced here. First, it is based on production and not export data. Second, the similarity between products is computed using the Product Space \citep{hidalgo2007} formula, while here we use the Sapling Similarity \citep{albora2023sapling}, which shows a higher prediction power in economic contexts. 

\section{Results}\label{sec:res}

\subsection{Regression Models}\label{sec:model}

\paragraph{Cross-sectional regression model}
We are primarily interested in the long‐run relationship between the covariates and the performance variables of interest. To filter out short‐term noise, we initially perform time averages of the dependent variables and covariates, thus moving from a panel data structure to a cross-sectional structure. In this structure, we fix a measurement time $t^*$, and regress the firm performance indicators (Growth $G$ and Labour Productivity) measured in period $[t^*,t^*+\Delta t]$ against the economic complexity variables and controls measured in period $[t^*,t^*-\Delta t]$.
Specifically, we estimate the following cross-sectional regression model:
\begin{equation}
\label{baseOLSmodel}
    Y_{f,t^*+\Delta t} = \beta X_{f,t^*} + \delta_{s} + \mu_{r} + \eta_{f,t^*}
\end{equation}
where $Y_{f,t^*+\Delta t}$ denotes firm performance (either value added growth or labour productivity) measured in $[t^*,t^*+\Delta t]$; $X_{f,t^*}$ denotes the vector of all the independent variables measured in $[t^*,t^*-\Delta t]$; and $\eta_{f,t^*}$ denotes the error term. To account for sectoral and geographical heterogeneity, $\delta_{s}$ comprises a set of 70 sectoral fixed effects based on the 2-digit ATECO classification (the Italian standard industrial classification, maintained by ISTAT and aligned with the EU NACE Rev.\ 2 taxonomy), with each firm assigned to the 2-digit ATECO sector reported in the ASIA financial dataset; $\mu_{r}$ comprises fixed effects for the Italian region in which the firm is located, accounting for the well-documented economic disparity between the northern and southern regions of Italy \citep{sbardella2021behind}. We detail in the following the variables $X_{f,t^*}$ that enter the model. We fix here $t^*$ equal to year 2015 and $\Delta t = 4$. 

%\msmc{I would introduce here the equation for the model with firm fixed effect, refer to it as ``panel regression model"} \textcolor{green}{siccome la definizione del modello cambia come cambiano anche le dependent variables, forse è necessario fare un paragrafo separato per cross e panel. Al posto del paragrafo FE metterei Panel regression model dove diciamo che le stesse dependent variables sopra sono year by year e il fatto dei 3 modelli con 3 lag differenti. }

\paragraph{Dependent variables} 
We separately examine Growth and Labour Productivity as the dependent variables. 
As these performance metrics typically follow fat-tailed distributions \citep{ishikawa2022statistical}, we work throughout
with log-transformed variables.
%To measure firm growth, we first smooth each firm’s value added using a backward-looking three-year average: $\overline{VA}_{f,t}=\text{mean}\{VA_{f,t-2},VA_{f,t-1},VA_{f,t}\}$.
%We then construct the growth variable as the difference between the logarithm of the smoothed value added in year $t+\Delta t$ and that at year $t$: \begin{equation}
%G_{f,t}=\log{\Biggl(\frac{\overline{VA}_{f,t+\Delta t}}{\overline{VA}_{f,t}}\Biggr)  }  
%\end{equation}
To measure firm growth, we first compute, for each firm, the average of the log value added over a three-year window:
\begin{equation}
\overline{\log VA}_{f,t} = \frac{1}{3}\sum_{s=t-2}^{t} \log VA_{f,s},
\end{equation}
where $VA_{f,s}$ is floored at a small positive value prior to taking logs, and the resulting average is equivalent to the log of the geometric mean of the value added over the window. We adopt this log-averaging approach, rather than averaging value added in levels, for consistency with the log-transformation used throughout to address the fat-tailed nature of firm-level performance variables \citep{ishikawa2022statistical}: since firm size is well approximated by a lognormal distribution \citep{sutton1997gibrat}, the arithmetic mean of value added in levels is disproportionately influenced by a small number of extreme observations, whereas the mean of the logs provides a more robust measure of central tendency. We then construct the growth variable as the difference between the smoothed log value added at $t+\Delta t$ and at $t$:
\begin{equation}
G_{f,t} = \overline{\log VA}_{f,t+\Delta t} - \overline{\log VA}_{f,t}.
\end{equation}
%We choose $\Delta t=4$. Similarly to growth, 
Labour productivity is defined as the corrisponding log difference with employment, $LP_{f,t} = \log VA_{f,t} - \log(\text{Employees}_{f,t})$, and is likewise measured as a three-year average over the same windows as the independent and dependent variables. As a control variable (initial labour productivity), it is averaged over 2015–2017, together with the other explanatory variables; as a dependent variable (future labour productivity), it is averaged over 2019–2021, matching the horizon used for the growth variable. Value added growth is accordingly computed as the log-difference between the smoothed value added in 2019–2021 and in 2015–2017.

%\textcolor{red}{check} 
%The labour productivity was originally computed on an annual basis. To smooth out fluctuations, we used its three-year average in the regression model. \textcolor{red}{check} So for instance, labour productivity, which serves as the dependent variable, is measured as the average over the years 2019–2021, while Value Added growth is computed as the difference between value added in 2019 and value added in 2021.
%In the panel specification, instead, we considered the 2015-2021 time span, and growth values are computed on 1, 2, and 3-year time horizons.

\paragraph{Independent variables}
We consider the four economic complexity indicators defined above as the main explanatory variables: EXPY; in-block diversification; out-of-block diversification; coherence. In agreement with the literature \citep{coad2009growth,sutton1997gibrat}, we include a size indicator (the firm's number of employees) as a control variable, together with firm age and initial labour productivity.
%Although each independent variable was originally computed on an annual basis, to smooth out fluctuations, we used backward-looking three-year averages in the regression model.\textcolor{red}{check}
%For example, if $C_{f,t}$ denotes the coherence of the firm $f$ at year $t$, then the corresponding covariate in the regression model is the backward-looking three-year average: $\overline{C_{f,t}} = \text{mean}\{C_{f,t-2},C_{f,t-1},C_{f,t}\}$.
%Since all covariates, except EXPY, span several orders of magnitude, we apply a logarithmic transformation to them before including them in the model.
%All explanatory variables are constructed as averages over the years 2015–2017. 
Although each independent variable was originally computed on an annual basis, we average it over a fixed three-year window to smooth out short-term fluctuations. For example, if $C_{f,t}$ denotes
the coherence of firm $f$ in year $t$, the corresponding covariate in the regression model is
\begin{equation}
\overline{C}_f = \frac{1}{3}\sum_{t=2015}^{2017} C_{f,t}.
\end{equation}
Since all covariates, except EXPY, span several orders of magnitude, we apply a logarithmic transformation to each variable prior to averaging, consistently with the approach adopted for the dependent variables. The resulting covariate is therefore the three-year average of the log-transformed variable. All explanatory variables are constructed as averages over the years 2015–2017.

\paragraph{Panel regression model}

While the cross-sectional analysis examines differences \textit{across} firms, we also investigate whether changes in export-basket characteristics \textit{within} firms are associated with subsequent growth. To this end, we complement the cross-sectional analysis with a panel specification covering the 2015--2021 period. We regress, for each year $t$, growth realized over the preceding $k$-year window on the same set of covariates measured at the start of that window, and include firm and year fixed effects:
\begin{equation}
\label{panelmodel}
    G_{f,t} = \beta X_{f,t-k} + \alpha_f + \gamma_t + \varepsilon_{f,t}, \qquad k = 1,2,3,
\end{equation}
where $G_{f,t} = \log VA_{f,t} - \log VA_{f,t-k}$ is the growth of value added realized between $t-k$ and $t$; $X_{f,t-k}$ is the vector of covariates measured $k$ years earlier, at $t-k$ (EXPY, in-block, out-of-block diversification, coherence, employees, and firm age, all log-transformed where relevant, as defined above); $\alpha_f$ is a firm ficed effect; $\gamma_t$ is a year fixed effect; and $\varepsilon_{f,t}$ is the error term clustered at the firm level. By lagging the covariates and re-estimating the model at each year $t$ over 2015--2021, this specification exploits substantially more of the panel's time variation than the cross-sectional model of Eq.~\eqref{baseOLSmodel}, which collapses the data into a single pair of pre- and post-period averages. Unlike the cross-sectional growth measure $G_{f,t^*}$, which is smoothed over three-year windows, the panel growth measure $G_{f,t}$ is not smoothed; we address its resulting sensitivity to extreme values separately in \ref{app:winsor}.
%Firm fixed effects account for time-invariant observed and unobserved heterogeneity across firms, while year fixed effects absorb common macroeconomic shocks and temporal trends common to all firms in a given year.

\subsection{Cross-section regression results}
%regression from original submission
\begin{comment}
{\def\arraystretch{1.15}\tabcolsep=10pt
\begin{table}[h!]
\centering
\caption{\textbf{OLS regression: export basket and performance.} EXPY and out-of-block diversification have a positive association with growth, whereas in-block diversification is negatively associated. On the other hand, EXPY and coherence are both positively associated with Profit for Employee, whereas both the in- and the out-of-block diversification exhibit no significant association.}

\begin{tabular}{l c c}
\toprule
 & \textbf{Growth} & \textbf{Profit per Employee} \\
\midrule

log Operative Revenue & 0.039$^{***}$ & 0.352$^{***}$\\
 & (0.003) & (0.014)\\
log Coherence & 0.012 & 0.118$^{**}$ \\
 & (0.009) & (0.046)\\ 
EXPY & 0.053$^{***}$ & 0.153$^{***}$ \\
 & (0.007) & (0.033)\\
log Out-of-block Diversification & 0.016$^{***}$ & 0.023\\
 & (0.004) & (0.022)\\
log In-block Diversification & -0.014$^{***}$ & -0.003\\
 & (0.005) & (0.025)\\
\\
 Sector dummies & YES & YES \\
\midrule
Observations & 12852 & 12852 \\
Adjusted R-squared & 0.066 & 0.100 \\
\bottomrule
\end{tabular}\\
\medskip
\textit{Robust standard errors in parentheses (HC1). $^{***}p<0.01$, $^{**}p<0.05$, $^{*}p<0.1$.} 
\label{main_table}
\end{table}
}
\end{comment}

\begin{table}[htbp]
    \centering
    \caption{\textbf{Cross-sectional regression: export basket and growth of value added.} EXPY and
out-of-block diversification have a positive association with growth, whereas in-block
diversification is negatively associated. Coherence, on the contrary, exhibits no significant
association.}
    \label{main_table1}
\def\sym#1{\ifmmode^{#1}\else\(^{#1}\)\fi}
\begin{tabular}{l*{3}{c}}
\toprule
                    &\multicolumn{1}{c}{(1)}&\multicolumn{1}{c}{(2)}&\multicolumn{1}{c}{(3)}\\
                    &\multicolumn{1}{c}{VA Growth}&\multicolumn{1}{c}{VA Growth}&\multicolumn{1}{c}{VA Growth}\\
\midrule

EXPY     &       0.171\sym{***}&       0.158\sym{***}&       0.149\sym{***}\\
                    &     (0.013)         &     (0.013)         &     (0.013)         \\
[1em]
In-block diversification (log)    &                     &      -0.053\sym{***}&      -0.053\sym{***}\\
                    &                     &     (0.007)         &     (0.008)         \\
[1em]
Out-of-block diversification (log)    &                     &       0.024\sym{***}&       0.021\sym{***}\\
                    &                     &     (0.005)         &     (0.005)         \\
[1em]
Coherence (log) &                     &                     &      -0.007         \\
                &              &              &     (0.018)         \\
[1em]
Labour productivity (log)           &      -0.396\sym{***}&      -0.393\sym{***}&      -0.394\sym{***}\\
                    &     (0.013)         &     (0.013)         &     (0.013)         \\
[1em]
Employees (log)        &       0.134\sym{***}&       0.144\sym{***}&       0.142\sym{***}\\
                    &     (0.007)         &     (0.007)         &     (0.008)         \\
[1em]
Firm age (log)            &      -0.056\sym{***}&      -0.056\sym{***}&      -0.060\sym{***}\\
                    &     (0.009)         &     (0.009)         &     (0.009)         \\

                    &                     &                     &     (0.018)         \\
\\
Sector FE       & Yes & Yes & Yes  \\
Region FE      & No & Yes  & Yes   \\
\midrule
N                   &   57588         &   57588         &   57588                  \\
$R^2$                 &       0.104         &       0.105         &       0.107                 \\
Adj $R^2$               &       0.103         &       0.104         &       0.106                 \\
\bottomrule
\multicolumn{4}{l}{\footnotesize Robust standard errors in parentheses}\\
\multicolumn{4}{l}{\footnotesize \sym{*} \(p<0.10\), \sym{**} \(p<0.05\), \sym{***} \(p<0.01\)}\\
\end{tabular}
\end{table}

\begin{table}[htbp]
    \centering
    \caption{\textbf{Cross-sectional regression: export basket and future labour productivity.} EXPY and out-of-block diversification have a positive association with labour productivity, whereas in-block diversification is negatively associated. Coherence, on the contrary, exhibits no significant association. All explanatory variables are constructed as averages over the years 2015--2017, whereas labour productivity, which serves as the dependent variable, is measured as the average over the years 2019--2021.}
    \label{main_table2}
\def\sym#1{\ifmmode^{#1}\else\(^{#1}\)\fi}
\begin{tabular}{l*{3}{c}}
\toprule
                    &\multicolumn{1}{c}{(1)}&\multicolumn{1}{c}{(2)}&\multicolumn{1}{c}{(3)}\\
                    &\multicolumn{1}{c}{LP}&\multicolumn{1}{c}{LP}&\multicolumn{1}{c}{LP}\\
\midrule

EXPY      &       0.159\sym{***}&       0.149\sym{***}&       0.138\sym{***}\\
                    &     (0.012)         &     (0.012)         &     (0.012)         \\
[1em]
In-block diversification (log)     &                     &      -0.041\sym{***}&      -0.043\sym{***}\\
                    &                     &     (0.007)         &     (0.008)         \\
[1em]
Out-of-block diversification (log)    &                     &       0.016\sym{***}&       0.012\sym{**} \\
                    &                     &     (0.004)         &     (0.005)         \\
[1em]
Coherence (log) &                     &                     &      -0.010         \\
                    &                     &                     &     (0.017)         \\  
[1em]
Initial Labour Productivity             &       0.567\sym{***}&       0.569\sym{***}&       0.567\sym{***}\\
                    &     (0.013)         &     (0.013)         &     (0.013)         \\
[1em]
Employees (log)         &       0.147\sym{***}&       0.155\sym{***}&       0.155\sym{***}\\
                    &     (0.007)         &     (0.007)         &     (0.007)         \\
[1em]
Firm age (log)             &       0.014\sym{*}  &       0.014\sym{*}  &       0.009         \\
                    &     (0.008)         &     (0.008)         &     (0.008)         \\
\\
Sector FE       & Yes & Yes & Yes  \\
Region FE      & No & Yes  & Yes   \\
\midrule
N                   &   57588         &   57588         &   57588         \\
$R^2$                  &       0.292         &       0.292         &       0.294         \\
Adj $R^2$                &       0.291         &       0.292         &       0.293         \\
\bottomrule
\multicolumn{4}{l}{\footnotesize Robust standard errors in parentheses}\\
\multicolumn{4}{l}{\footnotesize \sym{*} \(p<0.10\), \sym{**} \(p<0.05\), \sym{***} \(p<0.01\)}\\
\end{tabular}
\end{table}

% \begin{figure}[h!]
% 	\centering
% 	\includegraphics[width=0.7\textwidth]{OPRE-EXPY_volume-GROWTH-filtered.png}
    
% 	\caption{\textbf{The EXPY-Operative Revenue plane and growth}. The plane is divided into a grid of squares, each representing a cell populated by firms. The color of the square is given by the average growth of the firms in the square; the square is white if no firm is present. We applied a smoothing procedure using the Gaussian filter from the \url{scipy} Python package with $\sigma = 3$ to improve visualization. The figure suggests that both EXPY and Operative Revenue are positively associated with firm growth. \textcolor{red}{aggiornare figure e descrizione}
%     }
% 	\label{fig_expy}
% \end{figure} 

\begin{figure}[h!]
	\centering
	\includegraphics[width=0.6\textwidth]{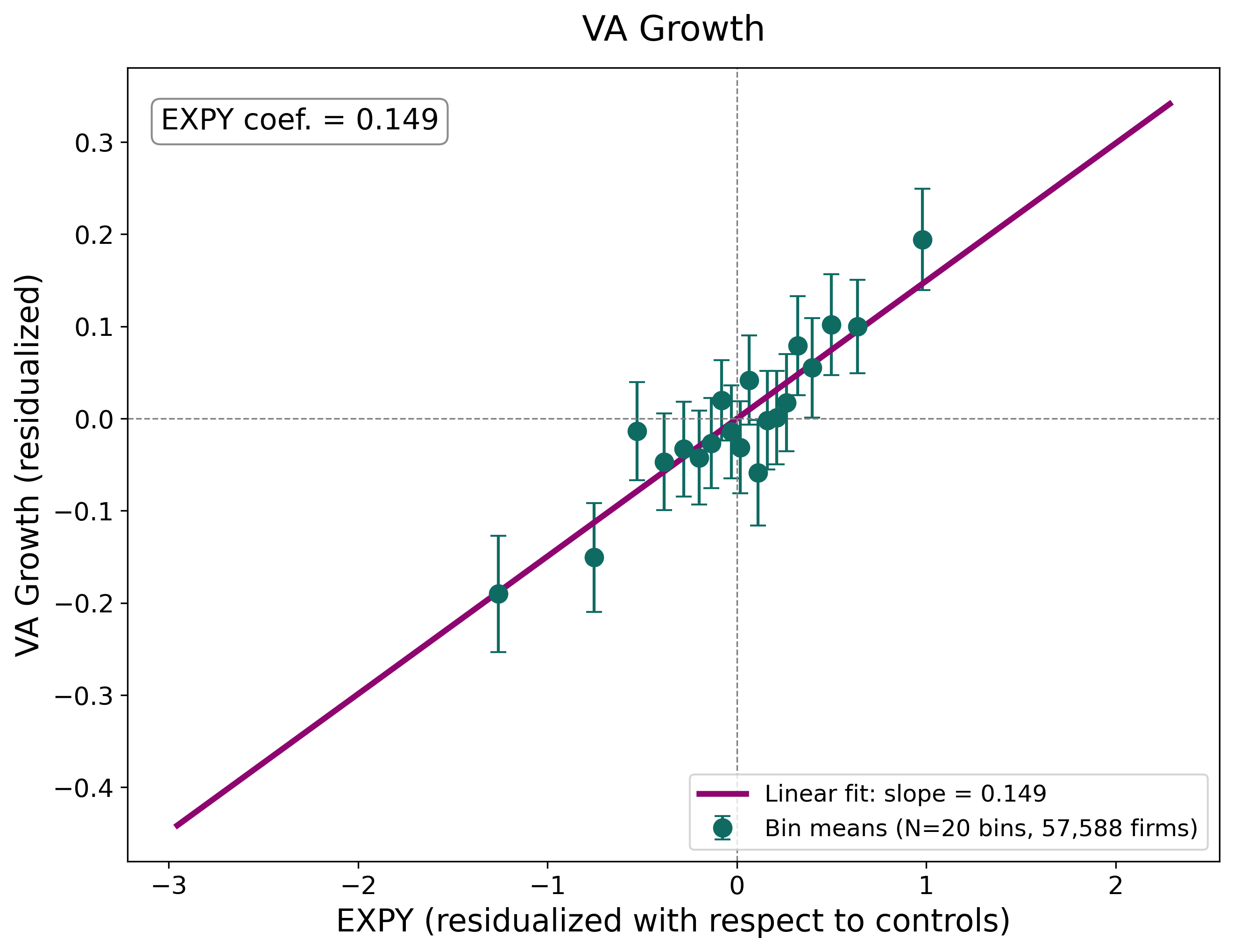}
    
	\caption{\textbf{Partial effect of EXPY on firm growth.} The panel show binned scatterplot of EXPY against VA Growth, after residualizing the variable with respect to initial productivity, in-block/out-of-block diversification, coherence, employees, firm age, and sector/region fixed effects (Frisch–Waugh–Lovell procedure). The slope of the fitted line corresponds exactly to the EXPY coefficient in the full regression (Table \ref{main_table1}).
    }
	\label{fig_expy}
\end{figure} 

In Tables \ref{main_table1} and \ref{main_table2} we illustrate the results of our cross-section regression analysis.
First, we find that \textit{EXPY is significantly associated with both future labour productivity and growth}. The coefficients are positive and significant in all regression models (Growth: $\hat{\beta} = 0.149, P<0.01$, Labour Productivity: $\hat{\beta} = 0.138, P <0.01$ in the most complete model). This finding allows us to go beyond previous works, which did not find a significant relationship between economic complexity indicators and firm growth.

Fig.\ref{fig_expy} illustrates this result graphically. Following the Frisch-Waugh-Lovell procedure, both VA Growth and EXPY are residualized with respect to all other covariates in the full model such as initial productivity, in-block and out-of-block diversification, coherence, firm size, firm age, and sector/region fixed effects, and firms are grouped into 20 equal-sized bins of residualized EXPY. The fitted line's slope reproduces exactly the EXPY coefficient in Table \ref{main_table1}, confirming that the positive association between EXPY and firm performance is not an artifact of firm size, initial productivity, or diversification patterns.

To interpret this result, we follow the line of reasoning of \cite{hausmann2007}, who show that EXPY predicts the growth of countries. On the one hand, they develop a dynamic model in which entrepreneurs face a cost-discovery process. In this respect, we are providing evidence in favor of a firm-level foundation of their indices. On the other hand, their indices are built \textit{ex post}: since high-income countries export those products, those products have to be, in some sense, of high quality or productivity. They confirmed this idea by empirically finding a positive relation between the EXPY of countries and growth. In principle, this reasoning can also be applied to firms -- and indeed we find that firms that tend to export high PRODY products experience higher growth as well. 
Finally, we report in \ref{sec:PRODY} a robustness test in which we use the standard PRODY measure instead of the skewness-aware logPRODY. The main results remain unchanged.

Our result connects the micro and macro levels by showing that the same concept and an analogous assessment of products' level of productivity play the same role both at the country and at the firm level.
Note that the same analogy does not hold for products' level of complexity, an assessment of the amount of country-level capabilities required to export a given product \citep{hidalgo2009,tacchella2012new,zaccaria2014,Angelini2017}.
Indeed, in \ref{sec:complexity}, we show that the connection between firms' performance and the average complexity of firms' exported products strongly depends on the specific definition of complexity. In particular, if the complexity value is computed (at the country level) using the Fitness approach \citep{tacchella2012new}, the effect is negative, while it is positive if the ECI/PCI method \citep{hidalgo2009} is adopted.
%We believe that this discrepancy might result from the substantially different algorithms proposed to measure the capability content of products
\begin{comment}
    \msmc{this is tautological and does not explain the result, and the following ``in any case" sounds dismissive. Shall we argue more why the effects are opposite? perhaps we could build the argument that PCI meaasures specialization and links to the income of the specializing countries (Mealy), while complexity the exclusivity of the product}.
\end{comment}

We believe this discrepancy reflects the different constructs the two indices capture. PCI is derived from a clustering of the country-product matrix that separates high-ECI countries (associated with high-income economies) from low-ECI ones \citep{mealy2019interpreting}; as such, it partly inherits an income-linkage similar in spirit to PRODY/EXPY (this is confirmed by the high correlation between PRODY and PCI). The Fitness-based Complexity, instead, captures products' weighted \textit{exclusivity} — how few, highly diversified countries export it — which need not coincide with income level or market demand. A firm exporting an exclusive-but-low-demand product may therefore score high on Complexity without benefiting from the market advantages that drive EXPY's positive effect. 

\begin{figure}[h!]
	\centering
	\includegraphics[width=\textwidth]{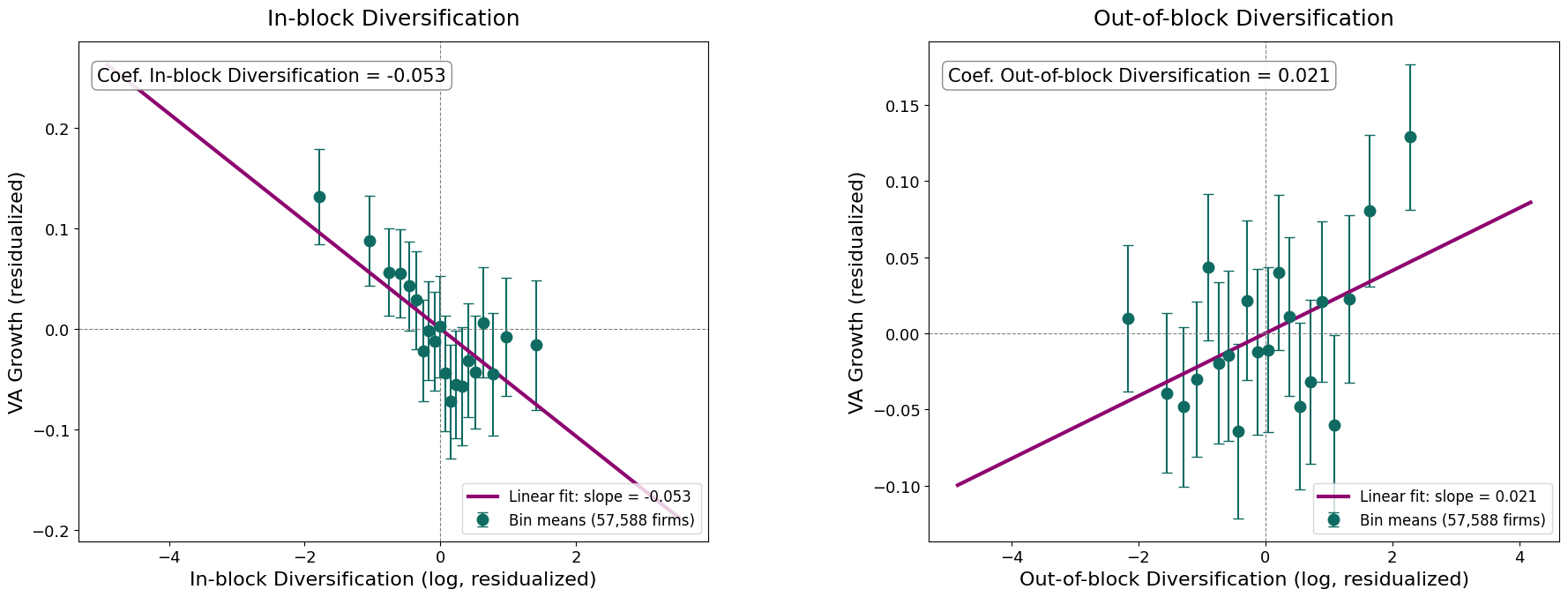}
    
	\caption{\textbf{Partial effects of in-block and out-of-block diversification on firm growth.}. Binned scatterplots of VA Growth against (left) in-block diversification and (right) out-of-block diversification (both in log), after residualizing all variables with respect to EXPY, initial labour productivity, coherence, firm size, firm age, and sector/region fixed effects (Frisch–Waugh–Lovell procedure). The slope of each fitted line corresponds exactly to the respective coefficient in the full multivariate regression (Table \ref{main_table1}, column 3): $\hat{\beta} = −0.053$ for in-block diversification and $\hat{\beta} =0.021$ for out-of-block diversification. The two panels illustrate the asymmetric role of diversification: expanding within the firm's core export block is associated with lower growth, while expanding beyond it is associated with higher growth.}
	\label{fig_block}
\end{figure}

Second, we find that \textit{the out-of-block diversification is positively associated with economic performance, whereas the in-block diversification exhibits a negative association} (in-block-diversification: $\hat{\beta} = -0.043 , P < 0.01$, out-of-block-div: $\hat{\beta} = 0.012 , P< 0.05$ in the most complete model to predict growth).
To further interpret the regression result, Fig.~\ref{fig_block} shows the partial relationship between each diversification indicator and firm growth, following the same Frisch–Waugh–Lovell residualization procedure used for Fig.~\ref{fig_expy}: both axes are residualized with respect to all other covariates in the full model (Table~\ref{main_table1}, column 3), so that the slope of the fitted line in each panel corresponds exactly to the respective regression coefficient. The left panel shows that, net of EXPY, initial productivity, coherence, firm size, firm age, and sector/region fixed effects, in-block diversification is negatively associated with growth (slope = $−0.053$): firms that expand their product range mostly within their core export block tend to grow less. The right panel shows the opposite pattern for out-of-block diversification (slope = $0.021$): firms that add products outside their core block tend to grow more, net of the same set of controls. Taken together, the two panels illustrate the asymmetric role of diversification documented in Table \ref{main_table1}: expanding within the firm's existing area of specialisation is associated with lower growth, while expanding among it is associated with higher growth.
% \begin{figure}[h!]
% 	\centering
% 	\begin{overpic}[width=0.47\textwidth]{GROWTH-VS-LOG-DIV_OVER_INBLOCK_v2.png}
% 		\put(3,88){\textbf{(A)}}
% 	\end{overpic}
% 	\hfill
% 	\begin{overpic}[width=0.47\textwidth]{INBLOCK-OUTBLOCK-GROWTH_filtered_newcolor.png}
% 		\put(3,88){\textbf{(B)}}
% 	\end{overpic}
% 	\caption{\textbf{Diversification patterns and growth.} \textbf{(A)} Results of a non-parametric regression for the average growth of firms versus the logarithm of the ratio between their total and in-block diversification. The result suggests that the higher the proportion of out-of-block exported products, the higher the growth. \textbf{(B)} An illustration of firm growth in the in-block vs. out-of-block diversification plane. We color each square by averaging the growth of firms at a fixed in-block and out-of-block diversification. The highest levels of growth are observed in the region with high out-of-block or low in-block diversification, in line with the regression results. \textcolor{red}{sostituire con nuove figure}}
% 	\label{fig_block}
% \end{figure}

We emphasize that this finding only holds once the blocks are identified through modularity maximization. A natural alternative would have been to use the Harmonized System classification (see www.wcotradetools.org/en/harmonized-system) to identify the blocks. Indeed, the World Customs Organization adopts a hierachical taxonomy to classify the exported products. The criteria include the possible natural origin, if the product is a raw material or a constituent material, the degree of processing, the function or use, etc. In this taxonomy, products are divided in 21 sections, which could be, in principle, used to partition the products in place of the community detection algorithm. In Table \ref{sector_div_table}) we repeat our regression analysis using the HS sections to define the corresponding out-of and in-section diversifications. While the in-section diversification maintains its negative effect, the out-of-section diversification is not significantly associated with growth or productivity. This result points to the essential role of network analysis in identifying the effects of different types of diversification on growth outcomes. 

{\def\arraystretch{1.15}\tabcolsep=10pt
\begin{table}[h!]
\centering
\caption{\textbf{Use of HS sections to define the blocks.} As shown in the previous tables, both in-block and out-of-block diversification measures are significantly associated with firms’ growth, where \textit{blocks} are defined from a data-driven community detection approach. By contrast, when blocks are measured using standard product classifications, which organize products in \textit{sections}, the corresponding out-of-section diversification shows no statistically significant association with growth nor future labour productivity.}
\label{sector_div_table}

\def\sym#1{\ifmmode^{#1}\else\(^{#1}\)\fi}
\begin{tabular}{l*{2}{c}}
\hline\hline
                    &\multicolumn{1}{c}{(1)}&\multicolumn{1}{c}{(2)}\\
                    &\multicolumn{1}{c}{VA Growth}&\multicolumn{1}{c}{LP}\\
\hline
EXPY     &       0.156\sym{***}&       0.142\sym{***}\\
                    &     (0.013)         &     (0.012)         \\
[1em]
In-section diversification (log)     &      -0.043\sym{***}&      -0.034\sym{***}\\
                    &     (0.009)         &     (0.008)         \\
[1em]
Out-of-section diversification (log)  &       0.007         &       0.003         \\
                    &     (0.005)         &     (0.005)         \\
[1em]
Coherence (log) &      -0.002         &      -0.003         \\
                    &     (0.019)         &     (0.017)         \\
[1em]
Labour Productivity (log)             &      -0.395\sym{***}&       0.567\sym{***}\\
                    &     (0.013)         &     (0.013)         \\
[1em]
Employees (log)         &       0.142\sym{***}&       0.154\sym{***}\\
                    &     (0.008)         &     (0.007)         \\
[1em]
Firm age (log)            &      -0.060\sym{***}&       0.009         \\
                    &     (0.009)         &     (0.008)         \\
\\
Sector FE       & Yes & Yes  \\
Region FE      & Yes & Yes   \\
\midrule
N                   &   57588         &   57588         \\
Adj $R^2$                &       0.105         &       0.293         \\
\hline\hline
\multicolumn{3}{l}{\footnotesize Robust standard errors in parentheses}\\
\multicolumn{3}{l}{\footnotesize \sym{*} \(p<0.10\), \sym{**} \(p<0.05\), \sym{***} \(p<0.01\)}\\
\end{tabular}
\end{table}
}
These results suggest that firm growth is not merely the result of idiosyncratic opportunities, but it can be anticipated by looking at whether the firm exhibits a high degree of diversification within and outside of its block. Based on the literature, the positive effect of the out-of-block diversification confirms the view that firms that diversify outside of their core block might possess transferable capabilities~\citep{matsusaka2001corporate,bernardo2002resources} which might allow them to benefit from opportunities in unrelated productive areas. 
On the other hand, the negative effect of the in-block diversification confirms the idea from learning theory~\citep{march1991exploration} that excessive focus
on related products may reflect a firm’s inability to benefit from opportunities outside of its core area, which can hinder growth.

\begin{comment}
pezzo vecchio di quando IND e OOBD non spiegavano profit per employee
Fourth, we find that \textit{neither the in-block nor the out-of-block diversification exhibits a significant association with the labour productivity} (in-block-div: $\hat{\beta} = -0.003 , P = 0.907$, out-of-block-div: $\hat{\beta} = 0.023 , P= 0.289$).
This suggests that most of the efficiency of a firm is captured by the coherence and EXPY metrics, with no benefits from looking at the diversification indicators.
Our interpretation of these findings is that while the EXPY captures whether a firm exports products that perform better in the market (and is therefore associated with both sales growth and high net income per employee), out-of-block diversification measures the extent to which a firm is expanding beyond its core activities. As such, it is positively correlated with Growth, but not with labour productivity, since this measure is net of costs. This can be explained by the fact that growing firms tend to invest heavily in new infrastructure, machinery, and technologies, which reduces their net earnings compared to larger, more established firms that invest less in expansion. \\
\end{comment}

Thus far, we have analyzed EXPY, as a measure of average product-level productivity, in-block diversification, and out-of-block diversification separately. A natural question emerging at this stage concerns the relationship among these three dimensions. As a first step, we examine whether out-of-block products are associated, on average, with higher PRODY values. To assess this hypothesis quantitatively, we perform a t-test comparing the mean PRODY values of in-block and out-of-block products. The difference is statistically significant ($t = 7.53, P < 0.001$) in the sample of 47,719 firms for which both in-block and out-of-block diversification can be measured. Although the magnitude of the difference is small in absolute terms, the pattern is systematic and statistically robust.

To assess whether this PRODY differential helps explain the growth effect of out-of-block diversification, we include these measures jointly in the regression as differentials and interaction terms. The complete results can be found in \ref{app:EXPY_INOUT}. We find that the PRODY differential between out-of-block and in-block products predicts firm growth, even after controlling for the firm’s overall EXPY level. This confirms that what matters is not simply exporting high-PRODY products, but specifically exporting out-of-block products whose PRODY is higher than that of the firm’s in-block products.

Finally, we find that \textit{Coherence does not exhibit a statistically significant association with labour productivity or growth}. This result aligns with the results of \cite{dosi2022firm}, who found no relation between their measure of coherence and growth. However, they claim
a positive relation between coherence and the relative profit margin of the firms, a quantity connected to, but different from, labour productivity.

\clearpage

\subsection{Panel regression}\label{panel_regression}

\begin{table}[htbp]
\centering
\caption{\textbf{Panel regression: export basket and value added growth.} The positive association between EXPY and growth in value added depends on the considered time interval, whereas in-block diversification is negatively associated at all time scales. Coherence exhibits no significant association.}
\def\sym#1{\ifmmode^{#1}\else\(^{#1}\)\fi}
\begin{tabular}{l*{3}{c}}
\hline\hline
  & \multicolumn{1}{c}{(1)}
  & \multicolumn{1}{c}{(2)}
  & \multicolumn{1}{c}{(3)} \\
  & \multicolumn{1}{c}{1 year}
  & \multicolumn{1}{c}{2 year}
  & \multicolumn{1}{c}{3 year} \\
\hline
EXPY
  & 0.030
  & 0.051\sym{**}
  & 0.081\sym{***} \\
  & (0.021)
  & (0.024)
  & (0.028) \\
[1em]
In-block diversification (log)
  & -0.048\sym{***}
  & -0.049\sym{***}
  & -0.060\sym{***} \\
  & (0.008)
  & (0.010)
  & (0.011) \\
[1em]
Out-of-block diversification (log)
  & -0.006
  & -0.008
  & 0.009 \\
  & (0.004)
  & (0.005)
  & (0.006) \\
[1em]
Coherence (log)
  & -0.017
  & -0.019
  & -0.010 \\
  & (0.017)
  & (0.020)
  & (0.022) \\
[1em]
Employees (log)
  & -0.549\sym{***}
  & -0.723\sym{***}
  & -0.848\sym{***} \\
  & (0.034)
  & (0.052)
  & (0.064) \\
[1em]
Firm age (log)
  & -0.184\sym{***}
  & -0.451\sym{***}
  & -0.561\sym{***} \\
  & (0.025)
  & (0.046)
  & (0.063) \\
[1em]
Constant
  & 2.235\sym{***}
  & 3.429\sym{***}
  & 4.071\sym{***} \\
  & (0.103)
  & (0.171)
  & (0.224) \\
\\
Firm FE
  & Yes
  & Yes
  & Yes \\
Year FE
  & Yes
  & Yes
  & Yes \\
\midrule
N
  & 345,528
  & 287,940
  & 230,352 \\
$R^2$
  & 0.057
  & 0.170
  & 0.316 \\
Within $R^2$
  & 0.003
  & 0.006
  & 0.007 \\
\hline\hline
\multicolumn{4}{l}{
  \footnotesize Standard errors clustered at firm level in parentheses
} \\
\multicolumn{4}{l}{
  \footnotesize
  \sym{*} \(p<0.10\),
  \sym{**} \(p<0.05\),
  \sym{***} \(p<0.01\)
} \\
\end{tabular}
\label{tab:va_growth_no_winsor}
\end{table}

Table \ref{tab:va_growth_no_winsor} shows the results of the panel regression analysis. The estimates show a positive association of EXPY with the growth of value added, which increases with the time horizon and is statistically significant at the 1\% level at the three-year horizon. 
This suggests that the estimated association between EXPY and growth becomes stronger over longer time horizons.
The panel results reveal an important distinction between the two diversification indicators. While the negative association with in-block diversification is observed both across and within firms and persists across all time horizons, the positive association with out-of-block diversification emerges primarily from comparisons across firms. Indeed, the within-firm association for out-of-block diversification is not statistically significant, except at the three-year horizon after winsorization\citep{ruppert2006trimming,wilcox2005trimming} (see \ref{app:winsor}).

Together, these findings suggest different roles for the two diversification indicators. They suggest that out-of-block diversification may serve as an indicator of firms' future growth prospects without necessarily being a driver of growth. However, the positive within-firm association observed at the three-year horizon after winsorization suggests that a medium-term within-firm relationship may also exist, although the results are sensitive to the treatment of extreme observations.
For in-block diversification, the consistent negative association across both specifications suggests a relationship that extends beyond persistent differences across firms. Establishing whether either relationship is causal remains an open question for future research.

We report three additional findings alongside our main results. First, weighted coherence, by contrast, is not statistically significant in any of the specifications. Second, we find a negative and highly significant relationship between firm size (the lagged logarithm of the number of employees) and growth of value added. This result is robust across all three time horizons. The magnitude of the effect increases progressively with the time horizon: in the winsorized specification shown in Table \ref{tab:va_growth_winsor}, the coefficient increases in absolute value from $-0.474$ on the one-year horizon to $-0.647$ on the three-year horizon. This pattern is consistent with a process of size convergence, whereby smaller firms tend to grow faster than larger ones—that is, mean reversion—in line with the literature on Gibrat’s law \citep{sutton1997gibrat}.
Finally, within $R^2$ remains relatively low across all specifications (0.003–0.007; see \ref{app:winsor} Table \ref{tab:va_growth_winsor}), indicating that the included regressors explain only a small proportion of within-firm variation in value-added growth. Such a result is not uncommon in this strand of the literature, as a substantial share of short-run growth variation is likely to be driven by idiosyncratic shocks that are not captured by structural fundamentals observed with a one-year lag. The more substantial overall $R^2$, which increases with the time horizon, instead suggests that the model’s overall explanatory power becomes stronger when medium-term growth dynamics are considered.

\clearpage

\section{Conclusions and discussion}\label{sec:conc}

When diversifying, corporate strategies investigate the risks and opportunities associated with the present and possible new products. Building on the theory of firm-level capabilities, previous studies show that firms tend to diversify in a way that is coherent with the present production basket, and coherence is positively associated with profit margin but not with growth. This evidence raises the question of finding product-level indicators that are able to capture corporate diversification strategies related to growth.
Such indicators exist at the country level. In particular, economic complexity indicators associate growth with measures of diversification and products' sophistication, reinterpreting the theory of capabilities as the building blocks of national economies. A precursor of these indicators is PRODY, which proxies products' productivity levels with the average income of the respective exporting countries. The average PRODY of the products exported by a country is called EXPY, and is positively associated with economic growth. 

In this paper, we investigate the export of more than 57,000 Italian firms and show that there is a systematic and robust association between their EXPY and both growth and labour productivity. The positive association between EXPY and growth at both country and firm levels suggests that similar underlying mechanisms may operate at the two scales. For example, exporting products typically exported by wealthier countries may reflect valuable productive capabilities at both the country and firm levels. Future research could place more emphasis on operationalizing the concept of capability on country-level and firm-level data, and detecting which capabilities are essential drivers of success for countries vs. firms.

We further examine whether the relationship between diversification and growth depends on where firms diversify relative to their core export areas. At the country level, diversification-based indicators are associated with growth. On the other hand, our findings indicate that at the firm level, one should distinguish between an in-block diversification (restricted products that belong to the export core of the firm and its competitors) and an out-of-block diversification (related to all the other products exported by the firm). Across firms, out-of-block diversification is positively associated with subsequent growth, whereas in-block diversification exhibits a negative association. The negative relationship with in-block diversification also emerges in the firm-fixed-effects analysis, while the positive out-of-block relationship is sensitive to the treatment of extreme observations

Our findings have various practical and theoretical implications. First, they call for a data-driven identification of production areas or blocks based on complex network methodologies and community detection, in particular. An incorrect identification of the blocks may lead to spurious findings, as shown by the mixed results obtained by using the Harmonized System classification to identify the blocks. From a theoretical perspective, our findings contribute to the literature on corporate coherence by pointing out a key difference between diversification associated with \textit{exploration capabilities}, which is associated to the ability to explore products outside the core production area and is associated with subsequent growth, and diversification associated with \textit{exploitation capabilities}, which is related to the ability to diversify in a coherent way and is negatively associated with growth both across and within firms. 
Note that well-defined blocks are only present at the firm level, and not at the country level \citep{laudati2023different}; therefore, the distinction between exploration and exploitation capabilities, if present, has to be subtler to detect at the country level, due to the non-modular structure of the country-product export matrix.

From a policy perspective, these findings offer valuable insights into how the level of productivity of exports (EXPY) and firm-level diversification strategies can anticipate economic performance. Policies that facilitate access to high-value markets and support strategic diversification could be informed by these findings. Relevant policy instruments could include flexible export promotion mechanisms, R\&D incentives, and the development of more accurate methods to map business ecosystems, such as network-based approaches. Such tools would enable policymakers to better understand the competitive environments of firms and identify opportunities for strategic expansion.
Overall, our results contribute to the discussion of firm growth and labor productivity, offering a perspective on how export-basket composition and diversification patterns are associated with firm economic performance.

\section{Acknowledgements}
%quando non saremo piu anonimi ringraziare Pietronero e Trombetti
We thank Andrea Mazzitelli for useful discussions.
M.S.M. acknowledges support from the Swiss National Science Foundation
(Grant No. 100013–207888).
A.Z. acknowledges the PRIN project No. 20223W2JKJ “WECARE”, CUP B53D23003880006, financed by the Italian Ministry of University and Research (MUR), Piano Nazionale Di Ripresa e Resilienza (PNRR), Missione 4 “Istruzione e Ricerca” - Componente C2 Investimento 1.1, funded by the European Union - NextGenerationEU.

\bibliographystyle{elsarticle-harv} 
\bibliography{cas-refs}
\clearpage
\appendix

\section{Sapling Similarity}\label{sec:sapling}
Starting from the seminal work of \cite{teece1994understanding}, extended at the macro level by \cite{hidalgo2007}, different measures of similarity between products or industries have been proposed in the literature \citep{hidalgo2018principle,neffke2013skill,zaccaria2014,cimini2022meta}. In this paper, we use the Sapling Similarity introduced by \cite{albora2023sapling}. The reason for this choice among the various possibilities lies in its predictive power. Various measures of relatedness or product similarity exist in the literature, but all share the same general idea that countries, on average, are more likely to export products that are similar to the ones they already export (but see also \cite{nomaler2024related}). So, one can compare different relatedness measures by checking their ability to predict new products \citep{albora2023product,gnecco2023matrix} and, in this sense, the Sapling Similarity outperforms the other similarity measures \citep{albora2023sapling}. In fact, the best methodologies for predicting new products are tree-based machine learning algorithms \citep{tacchella2023relatedness}, which, however, do not provide analytical similarity measures. Sapling Similarity incorporates some of the key elements that make machine learning a better predictor in a relatively simple mathematical formulation. In practice, Sapling Similarity $B_{pp'}$ is defined as a function of the co-occurrences $CO_{pp'}$ (how many firms export both $p$ and $p'$), the degrees $k_p$ and $k_{p'}$ (the number of firms that export product $p$ and $p'$, respectively) and the total number of firms in the bipartite network $N$:
\begin{equation}
	B_{pp'} =
	\begin{cases}
		1-f_{pp'} ~\text{ if } \frac{CO_{pp'}  N}{k_{p}k_{p'}}\geq1\\[3mm]
		-1+f_{pp'} ~\text{ otherwise, }
	\end{cases}	
\label{eq:sapling1}
\end{equation}
where:
\begin{equation}
f_{pp'} = \frac{CO_{pp'}\left(1-\frac{CO_{pp'}}{k_{p'}}\right)+\left(k_{p}-CO_{pp'}\right)\left(1-\frac{k_{p}-CO_{pp'}}{N-k_{p'}}\right)}
{k_{p}\left(1-\frac{k_{p}}{N}\right)}.
\label{eq:sapling2}
\end{equation}

The idea behind the derivation of these equations is to borrow some elements from tree-based machine learning algorithms, such as Random Forests and Boosted Trees, which show superior prediction performance compared to network-based models \citep{tacchella2023relatedness}. The starting point is a tree with just two leaves, reminiscent of a small decision tree, in which the goodness of the splitting is measured by using the Gini impurity. Using this simple tool, we can assess the similarity between $p$ and $p'$ by evaluating the influence of knowing that a product $p'$ is exported by a firm in our previous estimate of the probability that the same firm exports $p$. This influence is computed as the relative variation of the Gini impurity index between three cases: knowing that the generic firm $f$ exports $p$ (the bean), knowing that exports $p$ and $p’$, and knowing that exports $p$ but not $p'$ (the two leaves). If adding the information that $f$ exports $p’$ increases the probability that $f$ exports $p$, we will have a positive similarity. 

To clarify the above formula, we discuss two limit cases. Let us suppose that $CO_{pp'}$ (the number of co-occurrences) is equal to both the degrees $k_p$ and $k_{p'}$ (the ubiquities of product $p$ and $p'$, respectively). In this case, all firms exporting $p$ export also $p'$, so knowing that $f$ exports $p'$ implies that $f$ also exports $p$. This is mathematically translated into $f_{pp'}=0$ and Sapling Similarity $B_{pp'}=1$. In the opposite scenario, information about $p'$ does not vary our \textit{a priori} estimations about $p$. In formulas, this means that $k_{p}/N$ (the a priori probability that a generic firm exports $p$) is equal to $CO_{pp'}/k_{p'}$ (the probability that a firm exports $p$ restricted to the firms exporting $p'$). Imposing this condition in Eq.\ref{eq:sapling2} gives $f_{pp'}=1$ and a Sapling Similarity $B_{pp'}=0$. Note that this approach allows for a dependency on the total number of firms $N$ and the possibility of negative similarities (if knowing that a given firm exports $p'$ decreases our estimate that that firm exports $p$: this is expressed by \ref{eq:sapling1}), which are not considered in the usual economic complexity proximity measures. We refer the interested reader to the original publication \citep{albora2023sapling} for further details and the comparison among the out-of-sample prediction performances of different similarity measures.

\section{Community detection}

\subsection*{The BRIM algorithm}
\label{brim_section}
A graph can be considered modular if its nodes and edges form distinct clusters, or communities, where nodes within the same cluster are strongly connected, while the connections between nodes in different clusters are relatively sparse. A concrete example of a modular graph is in fact the firm-product graph, where firms of a certain industrial sector export (that is, are linked to) mainly products of the same sector. The degree of modularity is quantified using a metric called \textit{modularity}, which can be computed directly from the structural properties of the network. Let $\boldsymbol{A}$ be the adjacency matrix of the graph (that is, $A_{ij} = 1$ if node $i$ is linked to node $j$, $0$ otherwise),  $m$ be the total number of links and $k_i$ the degree of node $i$ (that is, the number of nodes connected to node $i$). Following \citep{newman2006modularity}, modularity $Q$ is defined as: 
\begin{equation}
    Q = \frac{1}{2m}\sum_{ij}\Biggl(A_{ij}-\frac{k_ik_j}{m}\Biggr)\,\delta(c_i,c_j),
\end{equation}
where $c_i$ denotes the community assignment of node $i$. Therefore, $\delta(c_i,c_j)=1$ if nodes $i$ and $j$ belong to the same community, $0$ otherwise. 
It follows that modularity is contingent not only on the graph structure but also on the specific node partition employed. A partition that yields high modularity indicates that most connections occur between nodes within the same community. A perfectly modular graph (that is, one with no links between nodes of different clusters) achieves $Q = 1$, while smaller modular structures approach zero. The BRIM algorithm \citep{barber2007} was specifically designed to maximize modularity in bipartite networks, that is, to find the partition $\{c_i\}$ such that Q is maximum.

\subsection*{Application of BRIM to our dataset}
\label{brim_applied}
We applied the BRIM algorithm to partition the firm-product bipartite graph into communities. However, since our dataset contains a separate firm-product graph for each year, running BRIM independently for each year results in partitions that vary slightly over time, which is undesirable for longitudinal analysis. To mitigate this problem, we restricted our analysis to firms that were present throughout the entire time span of our export data (2015–2021) and constructed a single aggregate firm-product graph, averaging connections throughout all years. To focus exclusively on products exported in tangible quantities, we employed the standard procedure utilized in Economic Complexity literature \citep{hidalgo2007, tacchella2012new}: we calculated the Revealed Comparative Advantage (RCA) for the resulting export matrix and eliminated all firm-product links with RCA values below $1$. The resulting graph is a non-weighted bipartite graph that links each firm to the products it exports, with an RCA greater than $1$.
The application of BRIM to this aggregated graph produced a partition with modularity $Q = 0.418$, corroborating previous findings \citep{laudati2023different}  that firm ecosystems exhibit a significant modular structure. Note that, as expected, the application of the same community detection algorithm using 6-digit or 4-digit products leads to slightly different blocks. Following the previous analyses of \cite{laudati2023different}, we use the 4-digit communities in this paper and checked that our results are robust if the 6-digit communities are used instead (not shown here).

\subsection*{Economic meaning of the blocks}

\begin{figure}[h!]
	\centering
	\includegraphics[width=\textwidth]{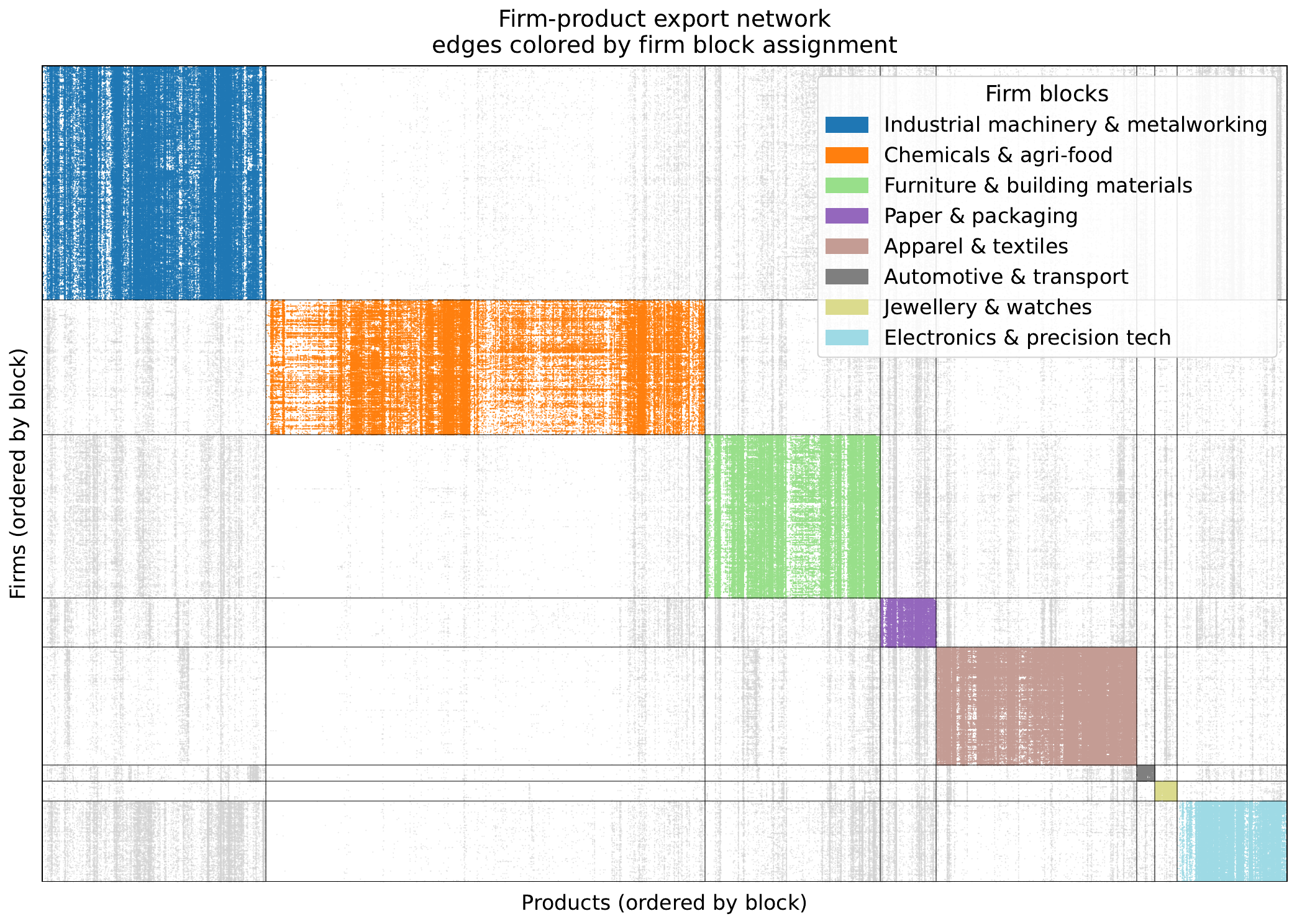}
	\caption{\textbf{Economic meaning of the blocks}. The figure represents the adjacency matrix of the bipartite network connecting firms (rows) to the products they export (columns). Rows and columns are arranged on the basis of the output of the community detection: firms and products that are assigned to the same block are put one next to the other. In such a way, the modular organization of the network is evident. Moreover, by inspecting the content of each block, we are able to assign an economic interpretation to each block. Each color corresponds to a different economic sector.}
	\label{FigBlocksEco}
\end{figure}

In Fig.\ref{FigBlocksEco} we illustrate the correspondence between the eight communities found by using the BRIM algorithm and economic sectors. By manually and automatically inspecting the products contained in each block, we are able to label each cluster by assigning an economic meaning to it. We remind that communities are identified by maximizing the modularity function; as such, products that are assigned to the same community are exported by the same firms (more precisely, the probability of being connected exceeds the one simply provided by their ubiquity). As a consequence, these communities identify also firms that export the same products, or products that belong to the same block. In Tables \ref{tab:firm_examples_2} and \ref{tab:firm_examples} we provide the list of products of two sample firms. In particular, we report which products belong to the same block of the firm and which products belong instead to a different block. Examining the description of products, one can easily see that while in-block products are highly similar and reflect the core activity of the firm, the out-of-block products represent a more explorative diversification pathway, in which the firm tries to discover new opportunities by expanding its exports with respect to the usual focus.

\begin{table}
\caption{Example of in-block (core) and out-of-block (outside core) products for a firm in the \textit{Furniture \& building materials} block.}
\label{tab:firm_examples_2}
\begin{tabular}{p{4cm}p{2cm}p{8cm}}
\toprule
Category & HS code & Product description \\
\midrule
In-block  & 271490 & Bitumen and asphalt, natural; asphaltites and asphaltic rock \\
In-block  & 321490 & Mastics; n.e.c. in heading no. 3214 \\
In-block  & 350699 & Glues and other adhesives; prepared, n.e.c. in heading no. 3506, not exceeding a net weight of 1kg \\
In-block  & 381600 & Refractory cements, mortars, concretes and similar compositions, including dolomite ramming mix, other than products of heading 38.01 \\
In-block  & 382450 & Mortars and concretes; non-refractory \\
Out-of-block  & 280421 & Gases, rare; argon \\
Out-of-block  & 280429 & Gases, rare; other than argon \\
Out-of-block  & 280430 & Nitrogen \\
Out-of-block  & 382000 & Anti-freezing preparations and prepared de-icing fluids \\
Out-of-block  & 391610 & Ethylene polymers; monofilament, of which any cross-sectional dimension exceeds 1mm, rods, sticks and profile shapes, whether or not surface-worked but not otherwise worked \\
\bottomrule
\end{tabular}
\end{table}

\begin{table}
\caption{Example of in-block (core) and out-of-block (outside core) products for a firm in the \textit{Agrifood and chemicals} block.}
\label{tab:firm_examples}
\begin{tabular}{p{4cm}p{2cm}p{8cm}}
\toprule
Category & HS code & Product description \\
\midrule
In-block  & 020130 & Meat; of bovine animals, boneless cuts, fresh or chilled \\
In-block  & 020312 & Meat; of swine, hams, shoulders and cuts thereof, with bone in, fresh or chilled \\
In-block  & 020319 & Meat; of swine, n.e.c. in item no. 0203.1, fresh or chilled \\
In-block  & 020422 & Meat; of sheep (including lamb), cuts with bone in (excluding carcasses and half-carcasses), fresh or chilled \\
In-block  & 020711 & Meat and edible offal; of fowls of the species Gallus domesticus, not cut in pieces, fresh or chilled \\
Out-of-block  & 391610 & Ethylene polymers; monofilament, of which any cross-sectional dimension exceeds 1mm, rods, sticks and profile shapes, whether or not surface-worked but not otherwise worked \\
Out-of-block  & 391733 & Plastics; tubes, pipes and hoses thereof, other than those of item no. 3917.31, not reinforced or otherwise combined with other materials, with fittings \\
Out-of-block  & 590900 & Textile hosepiping and similar textile tubing; with or without lining, armour or accessories of other materials \\
Out-of-block  & 680530 & Abrasive powder or grain; natural or artificial, on a base of materials n.e.c. in heading no. 6805, whether or not cut to shape or sewn or otherwise made up \\
Out-of-block  & 731990 & Sewing and knitting needles, bodkins, crochet hooks, embroidery stilettos and similar articles, for use in the hand, not elsewhere specified or included, of iron or steel \\
\bottomrule
\end{tabular}
\end{table}

\clearpage

\section{Average complexity of export baskets}
\label{sec:complexity}

As discussed in the main text, the PRODY index of a product $p$ is computed as the average GDP per capita of the countries that export $p$. As such, a high value of PRODY is associated with the sophistication, productivity, and market value of products. Various measures of product complexity are present in the economic complexity literature, trying to assess the capability content of the exported products. In Table \ref{expy_complexity_reg_table} we adopt the definitions by \cite{hidalgo2009} (Product Complexity Index, PCI) and \cite{tacchella2012new} (Complexity) and show that the association between the average complexity of the export basket of firms and economic performance depends on the definition of product complexity. In particular, the measure proposed by \cite{tacchella2012new} is negatively associated with economic performance. This would suggest that the level of sophistication and technological innovation of products is not straightforwardly connected to economic success, suggesting a more positive role for simpler but higher-demand products, as captured by the EXPY measure. However, PCI is positively associated with both growth and labour productivity. In order to explain this result, a starting point should be the interpretation of the economic complexity indexes provided by \cite{mealy2019interpreting}. In this paper, the author show that the ECI/PCI scheme is equivalent to performing a community detection algorithm to the bipartite country-product network, with the aim of finding two clusters of approximately equal size. As a result, the first cluster identifies high ECI countries, which are associated with high PCI products, and the second cluster identifies low ECI countries that export low PCI products. Whether this mechanism is able or not to quantify a sort of complexity is still subject of ongoing debate (see for instance \cite{mealy2019interpreting, Angelini2017,nomaler2022canonical,pietronero2017economic,du2026economic}); we leave a theoretical investigation of this behaviour for future work.
We point out that we performed a Variance Inflation Factor check (\cite{cohen2013applied}) to exclude the presence of collinearity among the explanatory variables.

{\def\arraystretch{1.15}\tabcolsep=10pt
\begin{table}[h!]
\centering
\caption{\textbf{Cross-sectional regression with EXPY and complexity assessments}. The average PCI of products is positively associated with economic performance indicators, while the contrary is true for Complexity. See text for more details.}
\begin{comment}
\begin{tabular}{l c c}
\toprule
 & \textbf{Growth} & \textbf{Profit per Employee}\\
\midrule

log Operative Revenue & 0.039$^{***}$ & 0.355$^{***}$\\
 & (0.003) & (0.013)\\
log Coherence & 0.012 & 0.121$^{***}$ \\
 & (0.009) & (0.046)\\ 
EXPY & 0.053$^{***}$ &  0.117$^{***}$\\
 & (0.008) & (0.039)\\
Average Complexity & 0.001 & 0.065$^{*}$\\
 & (0.007) & (0.037)\\
log Out-of-block Diversification & 0.016$^{***}$  & 0.020\\
 & (0.004) & (0.022)\\
log In-block Diversification & -0.014$^{***}$ & -0.004\\
 & (0.005) & (0.025)\\
 \\
Sector dummies & YES & YES \\
\midrule
Observations & 12852 & 12852 \\
Adjusted R-squared & 0.065 & 0.100\\
\bottomrule
\end{tabular}\\
\medskip
\textit{Robust standard errors in parentheses (HC1). $^{***}p<0.01$, $^{**}p<0.05$, $^{*}p<0.1$.}
\end{comment}
\def\sym#1{\ifmmode^{#1}\else\(^{#1}\)\fi}
\begin{tabular}{l*{2}{c}}
\hline\hline
                    &\multicolumn{1}{c}{(1)}&\multicolumn{1}{c}{(2)}\\
                    &\multicolumn{1}{c}{VA Growth}&\multicolumn{1}{c}{LP}\\
\hline
Average PCI (log)     &       0.057\sym{***}&       0.052\sym{***}\\
                    &     (0.008)         &     (0.007)         \\
[1em]
Average Complexity (log) &      -0.044\sym{***}&      -0.040\sym{***}\\
                    &     (0.010)         &     (0.009)         \\
[1em]
EXPY     &       0.106\sym{***}&       0.098\sym{***}\\
                    &     (0.014)         &     (0.014)         \\
[1em]
In-block diversification (log)      &      -0.044\sym{***}&      -0.034\sym{***}\\
                    &     (0.009)         &     (0.008)         \\
[1em]
Out-of-block diversification (log)    &       0.016\sym{***}&       0.008\sym{*}  \\
                    &     (0.005)         &     (0.005)         \\
[1em]
Coherence (log) &       0.023         &       0.017         \\
                    &     (0.018)         &     (0.017)         \\
[1em]
Initial Labour Productivity (log)                &      -0.398\sym{***}&       0.562\sym{***}\\
                    &     (0.014)         &     (0.013)         \\
[1em]
Employees (log)         &       0.138\sym{***}&       0.149\sym{***}\\
                    &     (0.008)         &     (0.007)         \\
[1em]
Firm age             &      -0.057\sym{***}&       0.010         \\
                    &     (0.009)         &     (0.008)         \\
\\
Sector        & Yes & Yes   \\
Region       & Yes & Yes     \\
\midrule
N                   &   53856         &   53856         \\
$R^2$                  &       0.110         &       0.290         \\
Adj $R^2$                &       0.108         &       0.289         \\
\hline\hline
\multicolumn{3}{l}{\footnotesize Standard errors in parentheses}\\
\multicolumn{3}{l}{\footnotesize \sym{*} \(p<0.10\), \sym{**} \(p<0.05\), \sym{***} \(p<0.01\)}\\
\end{tabular}

\label{expy_complexity_reg_table}
\end{table}
}

\clearpage

\section{PRODY robustness regressions}\label{sec:PRODY}

In this section, we report a robustness check concerning the measure of product sophistication and productivity used in this study, logPRODY \cite{Angelini2017}. This measure differs from that introduced by Hausmann et al. \cite{hausmann2007} in that the weighted average is computed using the logarithm of GDP per capita rather than the simple GDP per capita. The rationale for this transformation lies in the wide dispersion of GDP per capita values, which span several orders of magnitude. Applying the logarithmic transformation therefore makes countries with markedly different income levels more comparable \cite{Angelini2017}.

As a robustness check, we replicate the cross-sectional regressions presented in the main text, replacing the EXPY measure constructed using logPRODY with one constructed using the standard PRODY measure. The results clearly show that our findings are robust to this alternative specification of the EXPY measure.

\begin{table}[htbp]
    \centering
    \caption{\textbf{Cross-sectional regression: export basket and growth of value added.} EXPY (computed with simple GDP per capita) and out-of-block diversification have a positive association with growth, whereas in-block diversification is negatively associated. Coherence, on the contrary, exhibits no significant association.}
    \label{main_table1SI}
\def\sym#1{\ifmmode^{#1}\else\(^{#1}\)\fi}
\begin{tabular}{l*{3}{c}}
\toprule
                    &\multicolumn{1}{c}{(1)}&\multicolumn{1}{c}{(2)}&\multicolumn{1}{c}{(3)}\\
                    &\multicolumn{1}{c}{VA Growth}&\multicolumn{1}{c}{VA Growth}&\multicolumn{1}{c}{VA Growth}\\
\midrule

EXPY     &       0.150\sym{***}&       0.137\sym{***}&       0.130\sym{***}\\
                    &     (0.011)         &     (0.011)         &     (0.011)         \\
[1em]
In-block diversification (log)    &                     &      -0.055\sym{***}&      -0.055\sym{***}\\
                    &                     &     (0.007)         &     (0.008)         \\
[1em]
Out-of-block diversification (log)    &                     &       0.024\sym{***}&       0.021\sym{***}\\
                    &                     &     (0.005)         &     (0.005)         \\
[1em]
Coherence (log) &                     &                     &      -0.005         \\
                &              &              &     (0.018)         \\
[1em]
Labour productivity (log)           &      -0.396\sym{***}&      -0.393\sym{***}&      -0.394\sym{***}\\
                    &     (0.013)         &     (0.013)         &     (0.013)         \\
[1em]
Employees (log)        &       0.136\sym{***}&       0.146\sym{***}&       0.144\sym{***}\\
                    &     (0.007)         &     (0.007)         &     (0.008)         \\
[1em]
Firm age (log)            &      -0.056\sym{***}&      -0.055\sym{***}&      -0.060\sym{***}\\
                    &     (0.009)         &     (0.009)         &     (0.009)         \\

                    &                     &                     &     (0.018)         \\
\\
Sector FE       & Yes & Yes & Yes  \\
Region FE      & No & Yes  & Yes   \\
\midrule
N                   &   57586         &   57586         &   57586                  \\
$R^2$                 &       0.104         &       0.105         &       0.107                 \\
Adj $R^2$               &       0.103         &       0.104         &       0.105                \\
\bottomrule
\multicolumn{4}{l}{\footnotesize Robust standard errors in parentheses}\\
\multicolumn{4}{l}{\footnotesize \sym{*} \(p<0.10\), \sym{**} \(p<0.05\), \sym{***} \(p<0.01\)}\\
\end{tabular}
\end{table}

\begin{table}[htbp]
    \centering
    \caption{\textbf{Cross-sectional regression: export basket and future labour productivity.} EXPY (computed with simple GDP per capita) and out-of-block diversification have a positive association with labour productivity, whereas in-block diversification is negatively associated. Coherence, on the contrary, exhibits no significant association. All explanatory variables are constructed as averages over the years 2015–2017, whereas labour productivity, which serves as the dependent variable, is measured as the average over the years 2019–2021.}
    \label{main_table2SI}
\def\sym#1{\ifmmode^{#1}\else\(^{#1}\)\fi}
\begin{tabular}{l*{3}{c}}
\toprule
                    &\multicolumn{1}{c}{(1)}&\multicolumn{1}{c}{(2)}&\multicolumn{1}{c}{(3)}\\
                    &\multicolumn{1}{c}{LP}&\multicolumn{1}{c}{LP}&\multicolumn{1}{c}{LP}\\
\midrule

EXPY      &       0.135\sym{***}&       0.126\sym{***}&       0.117\sym{***}\\
                    &     (0.013)         &     (0.013)         &     (0.013)         \\
[1em]
In-block diversification (log)     &                     &      -0.043\sym{***}&      -0.044\sym{***}\\
                    &                     &     (0.007)         &     (0.008)         \\
[1em]
Out-of-block diversification (log)    &                     &       0.016\sym{***}&       0.013\sym{***} \\
                    &                     &     (0.004)         &     (0.005)         \\
[1em]
Coherence (log) &                     &                     &      -0.008         \\
                    &                     &                     &     (0.017)         \\
[1em]
Initial Labour Productivity             &       0.567\sym{***}&       0.569\sym{***}&       0.567\sym{***}\\
                    &     (0.013)         &     (0.013)         &     (0.013)         \\
[1em]
Employees (log)         &       0.149\sym{***}&       0.157\sym{***}&       0.157\sym{***}\\
                    &     (0.007)         &     (0.007)         &     (0.007)         \\
[1em]
Firm age (log)             &       0.014\sym{*}  &       0.014\sym{*}  &       0.009         \\
                    &     (0.008)         &     (0.008)         &     (0.008)         \\
\\
Sector FE       & Yes & Yes & Yes  \\
Region FE      & No & Yes  & Yes   \\
\midrule
N                   &   57586         &   57586         &   57586         \\
$R^2$                  &       0.292         &       0.292         &       0.294         \\
Adj $R^2$                &       0.291         &       0.292         &       0.293         \\
\bottomrule
\multicolumn{4}{l}{\footnotesize Robust standard errors in parentheses}\\
\multicolumn{4}{l}{\footnotesize \sym{*} \(p<0.10\), \sym{**} \(p<0.05\), \sym{***} \(p<0.01\)}\\
\end{tabular}
\end{table}

\clearpage

\section{VIF analysis}

In this section, we discuss potential correlations among the explanatory variables included in the regression models reported in Tables \ref{main_table1} and \ref{main_table2}. To assess this issue, we employ the Variance Inflation Factor (VIF) \cite{cohen2013applied}, which quantifies the extent to which the variance of an estimated regression coefficient abnormally increases because of the presence of correlations among the explanatory variables.\\ A threshold value of 5 is commonly used, with values above this level indicating potentially problematic multicollinearity and suggesting that the corresponding variable(s) may need to be excluded from the model. Conversely, VIF values close to 1 indicate the absence of multicollinearity. As shown in the tables, all the VIF values obtained in our analysis are close to 1, thus confirming that multicollinearity does not represent a concern in our regression models.

\begin{table}[htbp]
\centering
\caption{Variance Inflation Factors for the regressions in Table \ref{main_table1} and Table \ref{main_table2}}
\begin{tabular}{lcc}
\hline
Variable & VIF & 1/VIF \\
\hline
EXPY      & 1.11 & 0.903599 \\
In-block diversification           & 2.01 & 0.497770 \\
Out-of-block diversification           & 1.49 & 0.672527 \\
Coherence           & 2.10 & 0.477237 \\
Labour Productivity                    & 1.17 & 0.856941 \\
Employees          & 1.47 & 0.679683 \\
Firm age              & 1.15 & 0.866540 \\
\hline
Mean VIF & 1.50 & \\
\hline
\end{tabular}
\end{table}

\clearpage
\section{Integration between EXPY and diversification terms}\label{app:EXPY_INOUT}

\begin{table}[h!]
\centering
\caption{New variables: EXPY decomposition by block}
\label{tab:new_vars}
\begin{tabular}{lp{9cm}}
\toprule
\textbf{Variable} & \textbf{Description} \\
\midrule
\texttt{EXPY OF IBD} & Weighted average of z-score log PRODY across products exported by the firm that belong to its core production block (in-block products), weighted by export value. Missing if the firm exports no in-block products. \\
\addlinespace
\texttt{EXPY OF OOBD} & Weighted average of z-score log PRODY across products exported by the firm that fall outside its core production block (out-of-block products), weighted by export value. Missing if the firm exports no out-of-block products. \\
\addlinespace
\texttt{EXPY DIFF} & Difference between \texttt{EXPY OF OOBD} and \texttt{EXPY OF IBD}. Positive values indicate that a firm's out-of-block products are associated with higher productivity levels than its in-block products. \\
\addlinespace
\texttt{EXPY OF IBD (filled)} & Equal to \texttt{EXPY OF IBD}, with missing values replaced by zero.   \\
\addlinespace
\texttt{EXPY OF OOBD (filled)} & Equal to \texttt{EXPY OF OOBD}, with missing values replaced by zero. Used jointly with \texttt{HasOutblock} to preserve the full estimation sample. \\
\addlinespace
\texttt{HasOutblock} & Binary indicator equal to 1 if the firm exports at least one product outside its core production block, 0 otherwise. Captures the extensive margin of out-of-block diversification. \\
\bottomrule

\end{tabular}
    \label{tab:interact_descr}
\end{table}

\begin{table}[]
    \centering
    
{
\def\sym#1{\ifmmode^{#1}\else\(^{#1}\)\fi}
\begin{tabular}{l*{4}{c}}
\hline\hline
                    &\multicolumn{1}{c}{(1)}&\multicolumn{1}{c}{(2)}&\multicolumn{1}{c}{(3)}&\multicolumn{1}{c}{(4)}\\
                    &\multicolumn{1}{c}{VA Growth}&\multicolumn{1}{c}{VA Growth}&\multicolumn{1}{c}{VA Growth}&\multicolumn{1}{c}{VA Growth}\\
\hline
EXPY      &       0.149\sym{***}&                     &       0.192\sym{***}&                     \\
                    &     (0.013)         &                     &     (0.017)         &                     \\
[1em]
In-block diversification     &      -0.053\sym{***}&      -0.042\sym{***}&      -0.044\sym{***}&      -0.051\sym{***}\\
                    &     (0.008)         &     (0.009)         &     (0.009)         &     (0.008)         \\
[1em]
Out-of-block diversification     &       0.021\sym{***}&       0.012\sym{**} &       0.012\sym{**} &       0.010\sym{*}  \\
                    &     (0.005)         &     (0.006)         &     (0.006)         &     (0.006)         \\
[1em]
EXPY OF IBD        &                     &       0.128\sym{***}&                     &                     \\
                    &                     &     (0.014)         &                     &                     \\
[1em]
EXPY OF OOBD       &                     &       0.110\sym{***}&                     &                     \\
                    &                     &     (0.015)         &                     &                     \\
[1em]
EXPY DIFF           &                     &                     &       0.044\sym{***}&                     \\
                    &                     &                     &     (0.012)         &                     \\
[1em]
EXPY OF IBD(filled) &                     &                     &                     &       0.122\sym{***}\\
                    &                     &                     &                     &     (0.012)         \\
[1em]
EXPY OF OOBD(filled) &                     &                     &                     &       0.103\sym{***}\\
                    &                     &                     &                     &     (0.015)         \\
[1em]
HasOutblock        &                     &                     &                     &       0.057\sym{***}\\
                    &                     &                     &                     &     (0.020)         \\
[1em]
Coherence &      -0.007         &       0.007         &       0.003         &      -0.005         \\
                    &     (0.018)         &     (0.019)         &     (0.019)         &     (0.018)         \\
[1em]
Labour Productivity            &      -0.394\sym{***}&      -0.398\sym{***}&      -0.397\sym{***}&      -0.395\sym{***}\\
                    &     (0.013)         &     (0.015)         &     (0.015)         &     (0.013)         \\
[1em]
Employees         &       0.142\sym{***}&       0.131\sym{***}&       0.131\sym{***}&       0.139\sym{***}\\
                    &     (0.008)         &     (0.008)         &     (0.008)         &     (0.008)         \\
[1em]
Firm age             &      -0.060\sym{***}&      -0.057\sym{***}&      -0.056\sym{***}&      -0.059\sym{***}\\
                    &     (0.009)         &     (0.010)         &     (0.010)         &     (0.009)         \\
\\
Sector FE       & Yes & Yes & Yes  & Yes\\
Region FE      & Yes & Yes & Yes  & Yes \\
\midrule
N                   &   57586         &   47716         &   47716         &   57586         \\
Adj $R^2$                &       0.106         &       0.107         &       0.106         &       0.106         \\
\hline\hline
\multicolumn{5}{l}{\footnotesize Robust standard errors in parentheses}\\
\multicolumn{5}{l}{\footnotesize \sym{*} \(p<0.10\), \sym{**} \(p<0.05\), \sym{***} \(p<0.01\)}\\
\end{tabular}
}
    \caption{Regression results with EXPY variable decomposed by block.}
    \label{tab:interact}
\end{table}

In this section, we further investigate the interplay among EXPY, in-block diversification, and out-of-block diversification. To this end, we estimate a regression model that includes the variables already employed in the main text, together with a set of additional variables specifically constructed for this analysis. These newly introduced variables are described in Table \ref{tab:interact_descr}. \\The results of our regression analysis are reported in Table \ref{tab:interact}.

Column (1) reports the regression presented in Column (3) of Table \ref{main_table1}, which represents our baseline specification.\\
Model 2 (separate in-block and out-of-block EXPY measures):
Both coefficients are positive and statistically significant, with the coefficient on in-block EXPY (0.128) being slightly larger than that on out-of-block EXPY (0.110). This finding indicates that product sophistication matters both within and outside the firm’s core activities. However, the underlying mechanism does not appear to be solely that out-of-block products have higher PRODY values, since the average difference is only +0.024. A limitation of this model is the reduction in sample size: we lose approximately 10,000 firms with no out-of-block products, meaning that Models 1 and 2 are not directly comparable in terms of the number of observations.\\
Model 3 (overall EXPY and the EXPY differential):
The coefficient on the EXPY differential is positive and statistically significant (0.044***). Thus, the PRODY differential between out-of-block and in-block products predicts firm growth even after controlling for the firm’s overall EXPY level. This provides an interesting result: what matters is not simply exporting products with high PRODY values, but specifically exporting out-of-block products whose PRODY is higher than that of the firm’s in-block products.\\
Model 4 (filled measures and an indicator for out-of-block products, full sample):
The coefficient on the out-of-block indicator is positive and statistically significant (0.057***), indicating that the mere presence of products outside the firm’s core block predicts firm growth, irrespective of their PRODY values. Both the filled in-block EXPY coefficients (0.122) and the filled out-of-block EXPY coefficients (0.103) remain statistically significant. This is our preferred main specification because it retains the full sample of 57,586 firms and is therefore directly comparable with Model (1). \\

\section{Interaction between diversification strategies and age}\label{app:interaction_age}

In order to check the possible effects of firms’ age on the different diversification strategies, we have built a regression model in which we make these variables interact. The two interaction terms (Firm age × In-block diversification and Firm age × Out-of-block diversification ) are not statistically significant. This means that, holding all other variables included in the model constant, we find no evidence that the relationship between out-of-block diversification and firm performance differs between younger and older firms. The same is true for the in-block diversification.

\begin{table}[]
    \centering
{ \def\sym#1{\ifmmode^{#1}\else\(^{#1}\)\fi} \begin{tabular}{l*{2}{c}} \hline\hline & \multicolumn{1}{c}{(1)} & \multicolumn{1}{c}{(2)} \\ & \multicolumn{1}{c}{VA Growth} & \multicolumn{1}{c}{LP} \\ \hline  EXPY & 0.149\sym{***} & 0.137\sym{***} \\ & (0.013) & (0.012) \\ [1em] In-block diversification (log) & -0.090\sym{***} & -0.052\sym{**} \\ & (0.028) & (0.026) \\ [1em] Out-of-block diversification (log) & 0.018 & 0.013 \\ & (0.021) & (0.019) \\ [1em] Coherence (log) & -0.007 & -0.010 \\ & (0.018) & (0.017) \\ [1em] Firm age (log) & -0.103\sym{***} & -0.001 \\ & (0.027) & (0.025) \\ [1em] Firm age (log) $\times$ In-block diversification (log) & 0.013 & 0.003 \\ & (0.009) & (0.008) \\ [1em] Firm age (log) $\times$ Out-of-block diversification (log) & 0.001 & -0.000 \\ & (0.006) & (0.006) \\ [1em] Labour Productivity (log) & -0.394\sym{***} & 0.567\sym{***} \\ & (0.013) & (0.013) \\ [1em] Employees (log) & 0.141\sym{***} & 0.155\sym{***} \\ & (0.008) & (0.007)  \\ [1em] \\ Sector FE & Yes & Yes \\ Region FE & Yes & Yes \\ \midrule N & 57,586 & 57,586 \\ $R^2$ & 0.107 & 0.294 \\ Adj. $R^2$ & 0.106 & 0.293 \\ \hline\hline \multicolumn{3}{l}{ \footnotesize Standard errors in parentheses } \\ \multicolumn{3}{l}{ \footnotesize \sym{*} \(p<0.10\), \sym{**} \(p<0.05\), \sym{***} \(p<0.01\) } \\ \end{tabular} 
}
    \caption{Cross-sectional regression model with interaction between age and diversification terms.}
    \label{tab:interact_age}
\end{table}

\clearpage

\section{Panel regression and outliers}\label{app:winsor}

Winsorization \citep{wilcox2005trimming,ruppert2006trimming} is a statistical technique used to limit the influence of extreme values (outliers) without removing them from the sample. Values below the 1st percentile and above the 99th percentile are replaced with the corresponding threshold values, while the remainder of the distribution and the number of observations remains unchanged.

We apply winsorization to the dependent variable of our panel regression exercise, namely the value-added growth rate, because its construction as a log difference makes it particularly sensitive to extreme changes when the initial value added is very small or affected by accounting artefacts. If left untreated, such observations could distort the coefficient estimates by assigning disproportionate weight to a small number of anomalous values.

A comparison of the two specifications (with and without winsorization) shows that winsorization systematically reduces the magnitude of all the main coefficients without altering either their signs or their statistical significance. This suggests that the procedure effectively limits the influence of genuine outliers without distorting the underlying economic relationships. Model fit also improves in the winsorized specification, both in terms of the overall $R^2$, which increases from 0.316 to 0.365 at the three-year horizon and, albeit only marginally, in terms of the within $R^2$.
\begin{table}[htbp]\label{tab:va_growth_winsor}
\caption{Value Added Growth: No Winsor vs. Winsor}\label{tab:va_growth_winsor}

\centering
\def\sym#1{\ifmmode^{#1}\else\(^{#1}\)\fi}
\resizebox{\textwidth}{!}{%
\begin{tabular}{l*{6}{c}}
\toprule
 & \multicolumn{3}{c}{No Winsor}
 & \multicolumn{3}{c}{Winsor} \\
\cmidrule(lr){2-4}\cmidrule(lr){5-7}
 & 1 year & 2 year & 3 year
 & 1 year & 2 year & 3 year \\
\midrule
EXPY
 & 0.030
 & 0.051\sym{**}
 & 0.081\sym{***}
 & 0.030\sym{*}
 & 0.045\sym{**}
 & 0.068\sym{***} \\
 & (0.021)
 & (0.024)
 & (0.028)
 & (0.018)
 & (0.018)
 & (0.022) \\
[1em]
In-block diversification (log)
 & -0.048\sym{***}
 & -0.049\sym{***}
 & -0.060\sym{***}
 & -0.043\sym{***}
 & -0.032\sym{***}
 & -0.040\sym{***} \\
 & (0.008)
 & (0.010)
 & (0.011)
 & (0.007)
 & (0.007)
 & (0.008) \\
[1em]
Out-of-block diversification (log)
 & -0.006
 & -0.008
 & 0.009
 & -0.004
 & -0.005
 & 0.010\sym{**} \\
 & (0.004)
 & (0.005)
 & (0.006)
 & (0.004)
 & (0.004)
 & (0.005) \\
[1em]
Coherence (log)
 & -0.017
 & -0.019
 & -0.010
 & -0.017
 & -0.020
 & -0.002 \\
 & (0.017)
 & (0.020)
 & (0.022)
 & (0.014)
 & (0.015)
 & (0.017) \\
[1em]
Employees (log)
 & -0.549\sym{***}
 & -0.723\sym{***}
 & -0.848\sym{***}
 & -0.474\sym{***}
 & -0.514\sym{***}
 & -0.647\sym{***} \\
 & (0.034)
 & (0.052)
 & (0.064)
 & (0.027)
 & (0.033)
 & (0.042) \\
[1em]
Firm age (log)
 & -0.184\sym{***}
 & -0.451\sym{***}
 & -0.561\sym{***}
 & -0.178\sym{***}
 & -0.374\sym{***}
 & -0.437\sym{***} \\
 & (0.025)
 & (0.046)
 & (0.063)
 & (0.022)
 & (0.033)
 & (0.045) \\
[1em]
Constant
 & 2.235\sym{***}
 & 3.429\sym{***}
 & 4.071\sym{***}
 & 1.989\sym{***}
 & 2.518\sym{***}
 & 3.042\sym{***} \\
 & (0.103)
 & (0.171)
 & (0.224)
 & (0.085)
 & (0.112)
 & (0.151) \\
\midrule
N
 & 345,528
 & 287,940
 & 230,352
 & 345,528
 & 287,940
 & 230,352 \\
$R^2$
 & 0.057
 & 0.170
 & 0.316
 & 0.063
 & 0.231
 & 0.365 \\
Within $R^2$
 & 0.003
 & 0.006
 & 0.007
 & 0.004
 & 0.006
 & 0.007 \\
Firm FE
 & Yes & Yes & Yes
 & Yes & Yes & Yes \\
Year FE
 & Yes & Yes & Yes
 & Yes & Yes & Yes \\
\bottomrule
\end{tabular}%
}
\begin{minipage}{\textwidth}
\vspace{2pt}
\footnotesize
Standard errors clustered at firm level in parentheses.
\sym{*} \(p<0.10\), \sym{**} \(p<0.05\),
\sym{***} \(p<0.01\).
The dependent variable is winsorized at the 1st and 99th percentiles
in the ``Winsor'' columns.
\end{minipage}
\end{table}

\end{document}